\documentclass[usenatbib]{mn2e}
\usepackage{amsmath}
\usepackage{graphicx}
\usepackage[usenames]{color}
\usepackage{euscript}
\usepackage{amssymb}
\usepackage{journals}
\definecolor{grey}{rgb}{0.5,0.6,0.7}

\title[dE/dS0 Stellar Populations and Kinematics]{Evolution of dwarf
early-type galaxies I.\\ Spatially resolved stellar populations and internal
kinematics of Virgo cluster dE/dS0 galaxies.}
\author[I. Chilingarian]{Igor V. Chilingarian$^{1,2}$\thanks{E-mail:
    Igor.Chilingarian@obspm.fr}\\
$^{1}$Observatoire de Paris-Meudon, LERMA, UMR~8112, 61 Av. de
  l'Observatoire, 75014 Paris, France\\
$^{2}$Sternberg Astronomical Institute, Moscow State University, 13 Universitetsky prospect, 119992, Moscow, Russia
}
\begin{document}

\date{Accepted 2008 Dec 17. Received 2008 Dec 10; in original form 2008 Oct 16}

\pagerange{\pageref{firstpage}--\pageref{lastpage}} \pubyear{2008}

\maketitle

\label{firstpage}

\begin{abstract} 

Understanding the origin and evolution of dwarf early-type galaxies remains
an important open issue in modern astrophysics. Internal kinematics of a
galaxy contains signatures of violent phenomena which may have occurred,
e.g. mergers or tidal interactions, while stellar population keeps a fossil
record of the star formation history, therefore studying connection between
them becomes crucial for understanding galaxy evolution.  Here, in the first
paper of the series, we present the data on spatially resolved stellar
populations and internal kinematics for a large sample of dwarf elliptical
(dE) and lenticular (dS0) galaxies in the Virgo cluster. We obtained radial
velocities, velocity dispersions, stellar ages and metallicities out to 1--2
half-light radii by re-analysing already published long-slit and
integral-field spectroscopic datasets using the {\sc NBursts} full spectral
fitting technique. Surprisingly, bright representatives of the dE/dS0 class
($M_B = -18.0 \ldots -16.0$~mag) look very similar to intermediate-mass and
giant lenticulars and ellipticals: (1) their nuclear regions often harbour
young metal-rich stellar populations always associated with the drops in the
velocity dispersion profiles; (2) metallicity gradients in the main
discs/spheroids vary significantly from nearly flat profiles to
$-0.9$~dex~$r_e^{-1}$, i.e. somewhat 3 times steeper than for typical
bulges; (3) kinematically decoupled cores were discovered in 4 galaxies,
including two with very little, if any, large scale rotation. These results
suggest similarities in the evolutionary paths of dwarf and giant early-type
galaxies and call for reconsidering the role of major mergers in the dE/dS0
evolution.

\end{abstract}


\section{Introduction} 

Dwarf elliptical (dE) and lenticular (dS0) galaxies, low-luminosity ($M_B
> -18.0$~mag) stellar systems with early-type morphology, represent the
numerically dominant galaxy population in nearby galaxy clusters and groups
\citep{FB94}. In the cold dark matter models of hierarchical galaxy
formation they are considered as the building blocks of presently observed
stellar systems (e.g. \citealp{WF91}). Thus, understanding their formation
and evolution becomes one of the important questions of modern astrophysics.
Several dE/dS0 formation scenarii were proposed (see review and discussion
in \citealp{deRijcke+05}), but none of them is able to fully explain all
observational properties of these galaxies simultaneously. A possible
diversity of evolutionary scenarii has been suggested by \cite{vZBS04} and
later investigated by \cite{LGB08} to explain properties of different dE
subclasses, although \cite{BBCG08_1} insist on the ram pressure stripping to
be the unique channel of dE formation.

Dwarf and giant early-type galaxies form two distinct sequences in the
absolute magnitude ($M_B$) vs effective surface brightness ($\langle \mu
\rangle_e$) diagram (see e.g. \citealp{FB94}) joining around $M_B =
-18.0$~mag, which is often referred as a luminosity separating these two
classes (see \citealp{BBCG08} and \citealp{KFCB08} for the recent
discussions). Although, some arguments are provided (e.g. \citealp{GG03})
for the continuity of the dwarf--giant sequence on the $M_B$ vs $\langle
\mu_e \rangle$ plot, it is clear that the effective surface brightness of dE
galaxies correlates with their luminosity, or, in other words, less luminous
galaxies are fainter in terms of surface brightness, making their
observations a challenging task. One has to keep in mind that dE galaxies
usually lack of the interstellar medium (ISM), therefore their spectra
normally contain no emission lines. This explains why the first kinematical
study of dEs \citep{BN90} appeared more than a decade after the publication
of the absorption-line kinematics of giant elliptical galaxies (e.g.
\citealp{BC75}).

New debates addressing the dE formation appeared recently:
\cite{BBCG08_1,BBCG08} argue for the ram pressure stripping to be the only
scenario of dE formation, hence making these galaxies evolutionary different
from giant early-type systems and reproducing their locus on the $M_B$ vs
$\langle \mu \rangle_e$ plot; while \cite{JL08} succeeded to explain the
observed discontinuity of giant and dwarf ellipticals in the
size--luminosity relation by comparing a large homogeneous imaging dataset
to the predictions of semi-analytical models of elliptical galaxy formation,
thus suggesting their common origin. Comparing the structural properties of
dwarf galaxies to their internal kinematics and stellar populations
therefore becomes a question of major importance for understanding dE/dS0
formation and evolution.

The significant progress in the astronomical instrumentation and new
detectors having low readout noise became the cornerstones to the
accomplishments of several projects assessing spatially-resolved kinematics
of diffuse elliptical galaxies in different environmental conditions.
Kinematical profiles presented in
\cite{DRDZH01,DRDZH03,GGvdM02,GGvdM03,Pedraz+02,SP02,vZSH04} revealed a
diversity of the degree of rotational support and presence of
kinematically-decoupled components in some objects
\citep{DRDZH04,GGvdM05,PCSA05,Thomas+06}. Detailed studies of the
morphological properties of dEs revealed embedded structures such as stellar
discs, sometimes harbouring low-contrast spiral arms, or bars in many
galaxies \citep{JKB00,BBJ02,LGB06}. These discoveries strengthen the
evolutionary connection between dwarf ellipticals and discy late-type
dwarfs.

At the same time, studies of the stellar population properties of dwarf
elliptical galaxies still remain a challenge both for observations and for
the data analysis, because absorption line strengths, a traditional
spectroscopic indicator of stellar populations (see e.g. \citealp{WFGB94}
for the description of the Lick system) require high signal-to-noise ratios
in order to reach reasonable quality of age and metallicity estimates. Some
attempts were made to derive dE stellar populations from the narrow-band
photometry \citep{Rakos+01}. However, the calibrations used by the authors
were based on Galactic globular clusters, therefore an assumption about very
old ages was done, resulting in low metallicity estimates ($-1.4$ to
$-0.6$~dex). The integrated measurements of dE stellar populations from the
Lick indices were made later \citep{GGvdM03,vZBS04} and revealed populations
having intermediate ages of 3--5~Gyr, and considerably higher metallicities
around $-0.5$~dex. These results were confirmed and extended to a
larger sample of Virgo cluster dEs recently \citep{Michielsen+08}.

A new family of techniques for recovering stellar population properties
from integrated light spectra appeared recently
\citep{CidFernandes+05,OPLT06,CPSA07}. These methods are based on the direct
fitting of the the stellar population models against the observed spectra in
the pixel space. Since every pixel of an absorption-line galactic spectrum
bears information about the stellar content of a galaxy, these techniques are
potentially much more sensitive than the line-strength indices dealing with
the individual spectral features. Another advantage of the pixel space
fitting is an easy way of getting rid of residual cosmic ray hits and
regions of the spectra contaminated by emission lines by excluding these
regions from the fitting procedure.

The full spectral fitting using the {\sc NBursts} code (see preliminary
description in \citealp{Chilingarian06,CPSK07}) allowing to extract internal
kinematics simultaneously with the parametrized star formation history, was
successfully used in the studies of integrated kinematical and stellar
population properties of dE galaxies in the Fornax \citep{Michielsen+07} and
Abell~496 \citep{Chilingarian+08} clusters. The method was shown to be
stable and working well even for the data having low signal-to-noise ratios.

Up-to now, a published sample of dwarf and low-luminosity early-type
galaxies with the spatially-resolved stellar population information (apart
from Local Group dwarfs studied by means of colour-magnitude diagrams)
contains only seven objects, all of them observed with the Multi-Pupil Field
Spectrograph \citep{ADM01} at the 6-m Bolshoi Teleskop Al'tazimutal'nij
(BTA) of the Special Astrophysical Observatory of the Russian Academy of
Sciences (SAO RAS). The sample includes four dE/dS0s in the Virgo cluster
\citep{CPSA07,CSAP07} and three low-luminosity E/S0s in groups: NGC~770
\citep{PCSA05}, NGC~126 and NGC~130 \citep{CSAP08}. In all cases the
two-dimensional maps of kinematics and stellar population properties were
obtained using the {\sc NBursts} technique. Chemically and evolutionary
decoupled structures which are quite common in giant elliptical and
lenticular galaxies \citep{Sil06,Kuntschner+06} were detected in the central
regions of six low-luminosity galaxies.

Presently, the growing amount of high-quality optical spectroscopic data on
dwarf galaxies becomes available in the archives of major observatories. 
Therefore, we decided to re-analyse the existing and available spectroscopic
data for dE/dS0 galaxies using the full spectral fitting technique. The
project aims at studying the connection between internal kinematics and
stellar populations of dE galaxies and understanding the mechanisms of their
formation and evolution, primarily in the cluster environment.

In the first paper of the series we provide the kinematical and stellar
population profiles of Virgo cluster dE/dS0 galaxies obtained from the new
analysis of several published datasets carried out using the {\sc NBursts}
full spectral fitting technique \citep{CPSK07,CPSA07}. The thorough
discussion of the obtained results will be given in the next papers. The
paper is organized as follows: in Section 2 we describe the spectroscopic
data, data reduction procedures, and define a sample of galaxies, Section
3 provides details about the techniques used to analyse the data, Section 4
presents the results of this analysis, the conclusions are given in Section
5.

\section{The data and the sample}

In the course of our study we have used intermediate-resolution ground-based
spectroscopic data coming from various sources. The criteria we applied to
select the datasets were: (1) only published data with available advanced
calibrations such as stellar templates, twilight spectra, or well-defined
spectral resolution information and well documented spectrograph; (2)
coverage of the blue/green spectral region (4800\AA $< \lambda <$5400\AA)
containing prominent absorption features; (3) intermediate spectral
resolution not lower than $R = 1300$ to be able to constraint kinematics
simultaneously with stellar populations and tackle the degeneracy between
metallicity and velocity dispersion; (4) we gave preference to the data
having intermediate or high signal-to-noise ratio and spatially resolved
datasets (i.e. longslit or IFU spectroscopy). Hereafter we present the
data collections used in order of their importance for our study and briefly
discuss essential data reduction steps which had to be considered prior to
the analysis.

\subsection{Palomar dE Project: Long-Slit Spectroscopy}

Observations of 16 Virgo Cluster dwarf and low-luminosity early-type
galaxies were conducted with the Double Spectrograph at the 5-m Palomar
telescope during two observing runs in March 2001 and April 2002. The
reduced, sky subtracted flux calibrated datasets were kindly provided by
L.~Van~Zee (PI of the project) in fall 2004. The two wavelength ranges were
observed with the double beam spectrograph: 4800--5700\AA\ with the
resolving power $R \approx 2200$, and 8250--8900\AA\ having higher spectral
resolution ($R \approx 5400$). Integration times ranged from 1200~sec for
VCC~1075 to 12000~sec for VCC~1308. The 2~arcsec wide slit spanned 2~arcmin
and was placed along major axes of the observed galaxies (see Table~2 in
\citealp{vZSH04}). A number of radial velocity standards (F7--K2 giants) was
observed.

We refer to all the details regarding observations and data reduction to
\cite{vZSH04}, where thorough kinematical analysis of these data is
presented. In this study we make a direct use only of the blue beam spectra
centered on the Mg$b$ triplet; the red beam kinematics from \cite{vZSH04} is
used for comparison with the kinematical profiles we obtain.

We used the spectra of HD~136202, HD~182572, and HD~187691 obtained with the
same instrument in March and September 2001 and provided to us together with
the spectra of galaxies, to compute the variations of the spectral line
spread function (LSF) along the wavelength and assess the precision of the
data reduction. The analysis of the 10 spectra of HD~182572 obtained for
different positions of the star on the slit allowed us to assess the 
spatial variations of the spectral resolution.

We fit the stellar spectra against the high-resolution spectra of the
corresponding stars available in the ELODIE.3.1 \citep{PSKLB07} library in
five wavelength segments covering the spectral range of the blue setup of
the Double Spectrograph and overlapping by 20~per~cent. The fitting was done
using the penalized pixel fitting procedure \citep{CE04} returning $v$,
$\sigma$, $h_3$, and $h_4$ coefficients of the Gauss-Hermite parametrization
\citep{vdMF93}. The heliocentric radial velocities available from the
ELODIE.3.1 library have been subtracted from the fitting results. Then the
heliocentric corrections were computed for every individual spectrum using
the {\sc iraf noao.rv.rvcorrect} task and applied to the measurements. 

In case of perfect and unbiased wavelength calibration this procedure would
result in zero radial velocities, while non-zero values would measure the
systemic errors of the dispersion relation. Variations of $\sigma$, $h_3$,
and $h_4$ along the wavelength are caused by the distortions introduced by
the optical system of the spectrograph both, in the collimator and in the
camera. The mean value of the $h_4$ coefficient anti-correlates with the slit
width: wider slit resulting in the $\Pi$-shaped line profile corresponds to
negative $h_4$, as in e.g. VLT FLAMES/Giraffe \citep{Chilingarian+08}, where
the physical diameters of optical fibers are larger than the diffraction limit of
the collimator.

\begin{figure}
\includegraphics[width=\hsize]{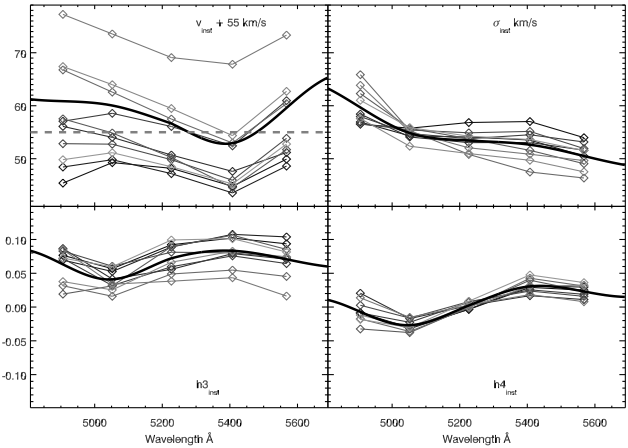}
\caption{Wavelength-dependent variations of the Gauss-Hermite
parametrization of the Double Spectrograph LSF at the Palomar 5-m Hale
Telescope obtained from the fitting of the 10 HD~182572 spectra. The adopted
curves are displayed by solid bold lines.\label{fig5mLSF}}
\end{figure}

In Fig~\ref{fig5mLSF} we present the variations of the Gauss-Hermite
parametrization coefficients along the wavelength obtained from the fitting
of 10 individual spectra of HD~182572. The adopted behaviour of the
coefficients which we use to transform the {\sc PEGASE.HR} models into the
resolution of the Double Spectrograph are shown by solid bold lines.

The measurements for individual spectra are scattered between $-10$ and
30~km~s$^{-1}$, which is probably caused by the flexures in the instrument
resulting in the overall shift of the wavelength solution zero-point.
Therefore, we can expect systematic errors of the radial velocity
measurements of an order of 20~km~s$^{-1}$. However, since our study is not
aimed at the accurate measurements of absolute radial velocities of
galaxies, this effect is not critical for us and affects none of our
scientific conclusions. We did not notice any correlation between the
measured systemic radial velocity offset and the position of the star on the
slit. Good agreement of the radial velocity profiles for some objects
with different datasets (see below) also suggests no important systematic
errors of the wavelength solution along the slit.

We also see an overall trend along the wavelength range, which is very well
reproduced by the measurements made on all individual spectra: the systemic
values first smoothly decrease by $\sim$7~km~s$^{-1}$ between 4800 and
5400\AA, and then increase by $\sim$8~km~s$^{-1}$ to the red end of the
wavelength range. This behaviour may be a result of an insufficient order of
the polynomial used to approximate the wavelength solution during the data
reduction procedure.

The spectral resolution ($\sigma_{\mbox{inst}}$) as a function of wavelength
(Fig~\ref{fig5mLSF}) rapidly rises from $\sim$65 to $\sim$55~km~s$^{-1}$
between 4800 and 5050\AA\ then slowly increasing to $\sim$48~km~s$^{-1}$ at
5650\AA. This behaviour keeps very similar for all the analysed spectra,
suggesting no important variations of the spectral resolution along the
slit.

Within the precision of the measurements ($\sim$0.03), the $h_3$ coefficient
remains stable along the wavelength range with the mean value of $+0.05$,
suggesting slight overall asymmetry of the DS blue arm LSF. The $h_4$
coefficient also demonstrates very stable behaviour for different wavelength
with the mean value of $0.00$ with a smooth gradient from $-0.03$ to
$+0.03$, which is the same order as the uncertainties of the measurements.

Hence, the overall variations of the Double Spectrograph LSF in the blue beam
exhibit very reproducible behaviour along the wavelength showing neither
significant changes along the slit, nor time-dependent drifts.  The
wavelength dependent variations of the spectral resolution and radial
velocity systematics are taken into account while fitting the spectra of
galaxies by convolving the {\sc PEGASE.HR} stellar population models using
the wavelength-dependent kernel, as described in Section~4.1 of
\cite{CPSA07}. Briefly, several (usually, five) sets of models are created
by convolving the original grid with LSF for different (five) positions
along the wavelength, then the linear interpolation between these sets is
done at every wavelength. This allows to obtain trustable velocity 
dispersion measurements down to $\sim$1/3 of the instrumental resolution 
(i.e., $\sim$20~km~s$^{-1}$) at intermediate signal-to-noise ratios
($\sim$20) which is very important for reliable estimations of the stellar
population parameters \citep{CCB08}.

The initial sample of \cite{vZSH04} included two galaxies having poor
signal-to-noise ratios: VCC~1743 and VCC~1857. For these objects we provide
only integrated measurements of kinematics and stellar populations along the
slit. For the other 14 objects we have performed the adaptive binning along
the slit in order to reach the minimal signal-to-noise ratio of 20 or 30 (10
for VCC~1075) per individual bin per pixel at 5200\AA. This was done by
co-adding individual consequent pixels along the slit from the centre
towards outer regions of the galaxy until the target signal-to-noise ratio
had been reached. The first bin was always assigned to the photometric
galaxy centre on the slit and grown in both directions. For the kinematical
and stellar population analysis the data have been rebinned logarithmically
in wavelength with the step corresponding to 40~km~s$^{-1}$.

\subsection{HyperLeda FITS Archive: Long-Slit Spectroscopy}

The HyperLeda project\footnote{http://leda.univ-lyon1.fr/}
\citep{Paturel+03} known mostly by its largest existing database of
homogenised structural, photometric, and kinematical parameters of galaxies,
also contains a rich collection of raw and processed spectral and imaging
datasets known as the HyperLeda FITS Archive (HFA). This resource includes
all the data on early-type galaxies presented in a series of papers by
Simien \& Prugniel originating from the CARELEC long-slit spectrograph at
the 1.93~m telescope of l'Observatoire de Haute Provence. In our study we
use the subset of this collection devoted to dwarf and low-luminosity
early-type galaxies in the Virgo cluster and presented in detail in
\cite{SP02}. The data were obtained during 5 observing runs in April and
June, 1999, February and March, 2000, and January, 2001. 
The CARELEC setup \#2 provided intermediate spectral resolution of $R
\approx 5000$ at the wavelength range $4700 < \lambda < 5600$\AA\ for the
5~arcmin long slit. The slit widths as specified in the metadata (FITS
headers) were: 2.2, 2.1, 2.3, 1.8, and 1.7 arcsec for the five observing
runs mentioned above. The integration times ranged from 60 minutes in a single
exposure for NGC~4476 to 210 minutes in four exposures for IC~3461. The
complete observing log is available in Table~2 of \cite{SP02}.

We used the original raw CCD frames available in the archive and reduced
them using our software package originally developed for the IFU data 
reduction and adapted to the long-slit spectroscopy. The main steps of the
data reduction include: (1) subtraction of bias; (2) cleaning cosmic ray
hits using Laplacian filtering \citep{vanDokkum01} (3) preliminary
flat-field correction using dome flat; (4) automatic identification of the 
arc line spectrum (He, Ne, Ar) and building the wavelength solution using the
3$^{\mbox{rd}}$ order polynomial; (5) rebinning the spectra logarithmically
in wavelength with the step of 25~km~s$^{-1}$; (6) residual flat field
correction using twilight frames. In average, 30 arc lines were used to build
the wavelength solution, resulting in the accuracy of about 0.04~\AA.

For all the observing runs, but April 1999, we used the twilight spectra
obtained in the same observing mode of the instrument in order to assess
variations of the CARELEC LSF along the wavelength range and along the slit.
We fit them against a high-resolution spectrum of the Sun available in the
ELODIE.3.1 spectral library exactly the same way as the stellar spectra from
the Palomar dE project. The LSF was shown not to exhibit any changes along
the slit. At the same time there are modest variations along the wavelength
range: the radial velocity systemic offset and $\sigma_{\mbox{inst}}$ (in
parenthesis) change smoothly from $+2$ (32 to 50 for different observing
runs) in the blue to $-2$ (25 to 40)~km~s$^{-1}$ in the red end of the
wavelength range; $h_3$ stays within 0.01 from zero; $h_4$ decreases from
$+0.04$ to $0.00$.

There are important changes of the instrumental resolution
$\sigma_{\mbox{inst}}$ between the observing runs because of different slit
widths used. No twilight spectra were obtained in the April 1999 observing
run, however, two stars included in the ELODIE.3.1 library were observed:
HD~84937 and HD~137759. The former is a relatively early type star (F5 III),
therefore its spectrum does not contain sufficient amount of deep spectral
lines to measure accurately the spectral resolution variations. However, the
latter one, a K-type giant, has an appropriate spectrum to determine the
spectrograph's LSF.

We did not use the data for four galaxies, IC~3120, IC~3457, IC~3653, and
NGC~4482. IC~3120 is a late-type dwarf and is out of the scope of this
study; IC~3653 was observed under very poor atmosphere conditions (6~arcsec
seeing); and the two remaining objects have too low signal (perhaps due to
poor transparency for NGC~4482) to determine stellar population parameters.
We notice that the data for IC~3457 also contain the spectrum of PGC~41516
(VCC~1399) on the slit, but it is also too faint even for integrated
measurements of stellar populations.

Finally, we used the data for 15 galaxies, 3 of those also having the data
available from the Palomar dE project. Spectra of some objects (e.g.
IC~783A) have poor signal-to-noise ratios, therefore we provide only
integrated measurements of the stellar population parameters along the slit.
For the remaining galaxies we performed the adaptive binning along the
slit to reach the minimal signal-to-noise ratio of 10 or 15 per bin per
pixel at 5200\AA\ using the technique described in the previous subsection.

\subsection{IFU Spectroscopy of Four dE Galaxies}

Four bright dE/dS0 galaxies, IC~783, IC~3468, IC~3509, and IC~3653 were
observed using the MPFS IFU spectrograph at the Russian 6-m telescope BTA
SAO RAS during 3 observing runs in March 2004, May 2004, and May 2005. The
details about observations, data reduction and analysis were presented in
\cite{Chilingarian06} and \cite{CPSA07,CSAP07}. All the data are available
in the ASPID archive \citep{ASPIDSR}.

Here we include the original measurements presented the corresponding papers
for two galaxies, VCC~490 and VCC~1871. We re-analysed the data for another
two objects, VCC~1422 and VCC~1545, improving the sky subtraction and
increasing the power of the multiplicative polynomial continuum to 15 (see
next section). The data were spatially rebinned using the Voronoi adaptive
binning technique \citep{CC03} to reach minimal target signal-to-noise ratio
per bin.

\subsection{SDSS DR6: Aperture Spectroscopy}

The Virgo cluster is partially contained in the footprint of the Sloan
Digital Sky Survey Data Release 6 \citep{SDSS_DR6}. Therefore, relatively
high signal-to-noise ratio flux calibrated intermediate-resolution ($R
\approx 1800$) spectra covering a large wavelength range ($3800 < \lambda <
9200$\AA) are available for many dE/dS0 cluster members. We use SDSS spectra
for the objects from the samples presented above in order to compare the
stellar population properties in the inner regions of the galaxies.
Relatively low spectral resolution of the data did not allow us to measure
velocity dispersions below 40~km~s$^{-1}$.

The spectra were obtained using the 2.5~m Apache Point Observatory telescope
with the multi-object double spectrograph containing 640 3-arcsec wide
circular fibres in the 3~deg field of view \citep{Gunn+06}. The
spectroscopic target selection of SDSS is based on the limiting magnitude
inside the fibre aperture, therefore the sample is biased towards nucleated
dwarf galaxies.

Since neither twilight spectra were available in the SDSS archive, nor
bright stars included in the ELODIE.3.1 library were targeted by the SDSS
due to high fluxes, it is not possible to directly measure the LSF
variations of the SDSS spectrograph along the wavelength. However, the
changes of the spectral resolution in terms of $\sigma_{\mbox{inst}}$ are
computed for each fibre by the SDSS data processing pipeline and provided
together with the data (in the 5$^{\mbox{th}}$ extension of FITS files
containing extracted spectra).  $\sigma_{\mbox{inst}}$ decreases from
68~km~s$^{-1}$ at 3800\AA\ to 52~km~s$^{-1}$ at 5800\AA. Then it sharply
jumps to 66~km~s$^{-1}$ smoothly decreasing down to 51~km~s$^{-1}$ at
9200\AA. This happens because SDSS spectra include two parts (blue and red)
obtained in the two different arms of the instrument. Here we fit the SDSS
spectra in the wavelength range between $3900 < \lambda <
6700$\AA~(rest-frame). In addition to a wide coverage of the spectral domain
resulting in better quality of the stellar population parameter estimates by
the full spectral fitting, the centre of the selected wavelength interval,
5300\AA, well corresponds to the discussed long-slit spectroscopic data.

23 of 31 galaxies included in our final sample have SDSS DR6 spectra
available.

Our final sample is presented in Table~\ref{tabsample}. We provide the
common galaxy designations, their numbers in the Virgo Cluster Catalogue
\citep{BST85} and availability of data in every of the discussed datasets.
The sample spans a range of luminosities $-18.27 < M_B < -15.10$~mag,
central velocity dispersions $23 < \sigma < 79$~km~s$^{-1}$. It is not a
statistically representative sample of the Virgo cluster dE galaxies,
because of the well-known correlation of the luminosity and effective
surface brightness of dE/dS0s, making fainter galaxies (in terms of the
luminosity) more difficult to observe. For several objects among the least
luminous members of our sample we provide only integrated measurements of
the stellar populations properties. The faintest object with the spatially
resolved information available is VCC~1308 ($M_B = -15.41$~mag).

\begin{table}

\caption{Final sample of low-luminosity dwarf early-type galaxies in the
Virgo cluster. Columns (1) and (2) give the object name and VCC number,
columns (3) -- (5) indicate availability of the data, column (6) provides
the source of photometric data. Galaxies observed 
with the MPFS IFU spectrograph are shown in italic.
\label{tabsample}
}
\begin{tabular}{lrcccc}
Name           & VCC  & Pal. & OHP & SDSS & Ph.ref$^{*}$ \\
\hline
\hline
IC~3081        &  178 & yes &  no & yes & 3 \\
IC~781         &  389 &  no & yes & yes & 4 \\
UGC~7399A      &  437 & yes &  no &  no & 1 \\
\emph{IC~783}  &  \emph{490} &  no & yes$^{1}$ & yes & 4 \\
UGC~7436       &  543 & yes & yes & yes & 1 \\
IC~783A        &  545 &  no & yes$^{1}$ & yes & 4 \\
NGC~4328       &  634 &  no & yes & yes & 4 \\
IC~3328        &  856 &  no & yes & yes & 1 \\
IC~3344        &  917 & yes &  no & yes & 2,3 \\
IC~3363        &  965 & yes &  no & yes & 3 \\
IC~3369        &  990 & yes &  no & yes & 3 \\
NGC~4431       & 1010 &  no & yes &  no & 4 \\
NGC~4436       & 1036 & yes & yes & yes & 2,3 \\
IC~3383        & 1075 & yes &  no & yes & 1 \\
IC~3381        & 1087 &  no & yes & yes & 1 \\
IC~3393        & 1122 & yes & yes & yes & 3 \\
IC~3413        & 1183 &  no & yes &  no & 4 \\
NGC~4476       & 1250 &  no & yes & yes & 1 \\
NGC~4482       & 1261 & yes &  no$^{2}$ & yes & 1 \\
IC~3437        & 1308 & yes &  no &  no & 3 \\
IC~3461        & 1407 &  no & yes$^{1}$ & yes & 1 \\
\emph{IC~3468} & \emph{1422} &  no &  no & yes & 1 \\
IC~3486        & 1491 &  no & yes & yes & 4 \\
PGC~41726      & 1514 & yes &  no &  no & 3 \\
\emph{IC~3509} & \emph{1545} &  no &  no & yes & 1 \\
IC~3602        & 1743 & yes$^{1}$ &  no & yes & 1 \\
IC~3647        & 1857 & yes$^{1}$ &  no &  no & 1 \\
\emph{IC~3653} & \emph{1871} &  no &  no$^{3}$ & yes & 1 \\
IC~3735        & 2019 & yes &  no &  no & 1 \\
IC~3773        & 2048 &  no & yes &  no & 1 \\
IC~3779        & 2050 & yes &  no & yes & 1 \\
\hline
\hline
\end{tabular}

\footnotesize{$^1$ poor signal-to-noise ratio, only integrated measurements
are possible}

\footnotesize{$^2$ very poor signal-to-noise ratio, object excluded}

\footnotesize{$^3$ poor seeing, object excluded}

\footnotesize{$^*$ Photometric references: 1 -- \cite{Ferrarese+06}, 2 --
\cite{Stiavelli+01} and/or \cite{GGvdM03}, 3 -- \cite{vZBS04}, 4 -- SDSS DR6}

\end{table}

\section{Data processing and analysis}

We have fit the high-resolution {\sc pegase.hr} \citep{LeBorgne+04}
simple stellar population (SSP) models against the observational data using
the {\sc NBursts} full spectral fitting technique \citep{CPSA07,CPSK07}. We
computed the SSP models for the \cite{Salpeter55} stellar initial mass
function using the recent version of the ELODIE stellar library (v~3.1,
\citealp{PSKLB07}) providing a spectral resolution of $R=10000$ in the
wavelength range $3900 < \lambda < 6800$~\AA. 

The spectral fitting method is based on the non-linear least square fitting
using the constrained Levenberg-Marquardt optimisation of the
following quantity:

\begin{eqnarray}
        \chi^2 = \sum_{N_{\lambda}}\frac{(F_{i}-P_{1p}(T_{i}(SFH) \otimes
        \mathcal{L}(v,\sigma,h_3,h_4) + P_{2q}) )^2}{\Delta F_{i}^2}, 
        \nonumber\\
        \mbox{where} \quad T_{i}(SFH) = \sum_{N_{bursts}}k_{i} T_{i}(t_n,
        Z_n)
\label{chi2eq}
\end{eqnarray}

\noindent
where $\mathcal{L}$ is the line-of-sight velocity distribution in the
Gauss-Hermite parametrization \citep{vdMF93}; $F_{i}$ and $\Delta F_{i}$ are
observed flux and its uncertainty; $T_{i}(SFH)$ is the flux from a synthetic
spectrum, represented by a linear combination of $N_{bursts}$ SSPs and
convolved according to the line-spread function of the spectrograph;
$P_{1p}$ and $P_{2q}$ are multiplicative and additive Legendre polynomials
of orders $p$ and $q$ for correcting the continuum; $t_n$ and $Z_n$ are age
and metallicity of the $n^{th}$ SSP. The relative contributions of the SSPs
and the additive continuum are fit linearly at every evaluation of the
function, whereas all the remaining parameters (including $P_{2q}$ for the
additive continuum) are derived from the the non-linear optimisation.
The needs for multiplicative continuum and possible side-effects have been
presented and analysed in detail in Appendix~A2.3 of \cite{CPSA07},
Appendix~B1 of \cite{Chilingarian+08}, and \cite{Koleva+08}.

If emission lines present in the spectrum, the wavelength regions
contaminated by them are excluded from the fitting. In the wavelength range
of SDSS we then exclude Balmer lines (7\AA-wide intervals), [OIII] ($\lambda
=$ 4363, 4959, 5007~\AA), [NII] ( $\lambda=$ 5199~\AA), [NI] ($\lambda =$
6548, 6583~\AA), and [SII] ($\lambda =$ 6717, 6731~\AA). Besides, we always
exclude from the fitting the spectral regions contaminated by the strong
air-glow lines: [OI] ($\lambda =$ 5577, 6300, 6364~\AA), NaI ($\lambda=$
5890, 5896\AA), and mercury lines originating from the light pollution, the
strongest one being HgI ($\lambda=$ 5461\AA).

The stellar population properties were derived in the parameter space with
the rotated age and metallicity axes as described in
Section~3 of \cite{Chilingarian+08}: 

\begin{eqnarray}
    \eta = (3 Z + 2 \log_{10} t) / \sqrt{13}; \nonumber\\
    \theta = (-2 Z + 3 \log_{10} t) / \sqrt{13}
\label{coordeq}
\end{eqnarray}

\noindent 
In this coordinate system the $\eta$ axis is parallel to the direction of
the well-known age--metallicity degeneracy for old stellar populations
in the age--metallicity space expressed by \cite{Worthey94} as ``if the
percentage change $\Delta \mbox{age} / \Delta Z \sim 3/2$ for two
populations, they will appear almost identical in most indices''.

All the results presented in this paper correspond to the fitting of a
single SSP. We did not include the additive polynomial in the fitting. The
adopted values of the multiplicative polynomial continuum were 13, 13, and
15 for the Palomar dE, HFA, and SDSS datasets respectively. In one of the
next papers of the series we will present the fitting of the nuclear regions
of the galaxies with a more complex model containing two stellar populations.

It was shown by \cite{Koleva+08} that the full spectral fitting using the
{\sc NBursts} code produced consistent stellar population parameters with
those derived from the measurements of the Lick indices, being several times
more precise. Here we also stress that (1) age determinations from the full
spectral fitting are insensitive to the masking of Balmer lines (see
Appendix~A2 in \citealp{CPSA07} and Appendix~B in
\citealp{Chilingarian+08}); (2) velocity dispersions can be precisely
determined at down to 1/3--1/2 of the spectral resolution (i.e.
15--20~km~s$^{-1}$ for the HFA data and 18--25~km~s$^{-1}$ for the Palomar
dE data) and these measurements remain unbiased (see e.g. \citealp{KBCP07}
and Section~2 in \citealp{CCB08}); (3) more optimal usage of the information
contained in the spectrum by the full spectral fitting compared to the
techniques using line-strength indices results in considerably higher
precision of the derived stellar population parameters or, reciprocally,
lower values of the required signal-to-noise ratio per \AA\ to get the
comparable uncertainties (see e.g. \citealp{KBCP07}); (4)
luminosity-weighted values of ages and metallicities are insensitive to the
$\alpha$/Fe ratios of the populations being fit
\citep{Chilingarian+08,Koleva+08} even when using the {\sc PEGASE.HR} SSP
models not representative of metal-rich $\alpha$-enhanced populations since
they are based on the empirical stellar library, which contains stars only
from the Solar neighbourhood, where $\alpha$/Fe abundance ratios are
correlated with their metallicities.

We have conducted a numerical experiment to quantify the age- and
metallicity-sensitive information in the absorption-line spectra in order to
illustrate the points (1) and (3) mentioned above. Since the flux at every
wavelength contributes to the total value of $\chi^2$ and this contribution
depends on many parameters, a possible way to estimate the sensitivity of
the fitting procedure $S(\lambda, p_0, \dots, p_n)$ to a given parameter
$p_i$ at a given wavelength $\lambda$ is to compute the corresponding
partial derivative of the template grid accounting for the multiplicative
continuum variations. In our case the sensitivity to a stellar population
parameter $\eta$ (and similarly, $\theta$) at a point of the parameter space
$(t_0, Z_0, \sigma_0)$ will be expressed as:

\begin{align}
S(\lambda, \eta)&|_{(t_0, Z_0, \sigma_0)} = \nonumber\\
 & \frac{\partial}{\partial \eta} \chi^2\{P_{1p}(\lambda, \eta, \theta)|_{(t_0, Z_0)}
    (T(\lambda, t, Z) \otimes \mathcal{L}(\sigma_0)) \},
\label{senseq}
\end{align}

\noindent with the quantities defined in Eq.~\ref{chi2eq},~\ref{coordeq}
assuming SFH containing a single SSP. This means, that the squared fitting
residuals of a spectrum $(\eta_0 + \Delta \eta, \theta_0)$ against the model
$(\eta_0, \theta_0)$ at every pixel would reflect its contribution to the
overall $\chi^2$ when varying $\eta$, i.e. its sensitivity to, or, in other
words, the measure of information about a given stellar population
parameter.

Practically, the computation is straightforward and has been conducted as
follows. We have chosen six {\sc PEGASE.HR} ``reference SSPs'' representing
stellar populations with ages of 1.5, 5, and 12~Gyr and metallicities of
-1.0 and -0.3~dex. Then, for every of the six ``reference SSPs'', six models
have been constructed by convolving them with three Gaussians corresponding
to the velocity dispersions of 60, 120, and 180~km~s$^{-1}$ and adjusting
$\eta$ and $\theta$ parameters corresponding to their ages and metallicities
(see Eq.~\ref{coordeq}) by 0.05~dex. For example, for the (5~Gyr,
$-0.3$~dex) ``reference SSP'' the pairs were constructed using (5.33~Gyr,
$-0.258$~dex) and (5.50~Gyr, $-0.328$~dex) respectively. Later, these models
were fit against their ``reference SSPs'' using the {\sc PPXF} procedure by
\cite{CE04} with the 13th order multiplicative polynomial continuum. The
fitting residuals obtained by this procedure correspond to the quantity
defined in Eq.~\ref{senseq}.

\begin{figure}
\includegraphics[width=\hsize]{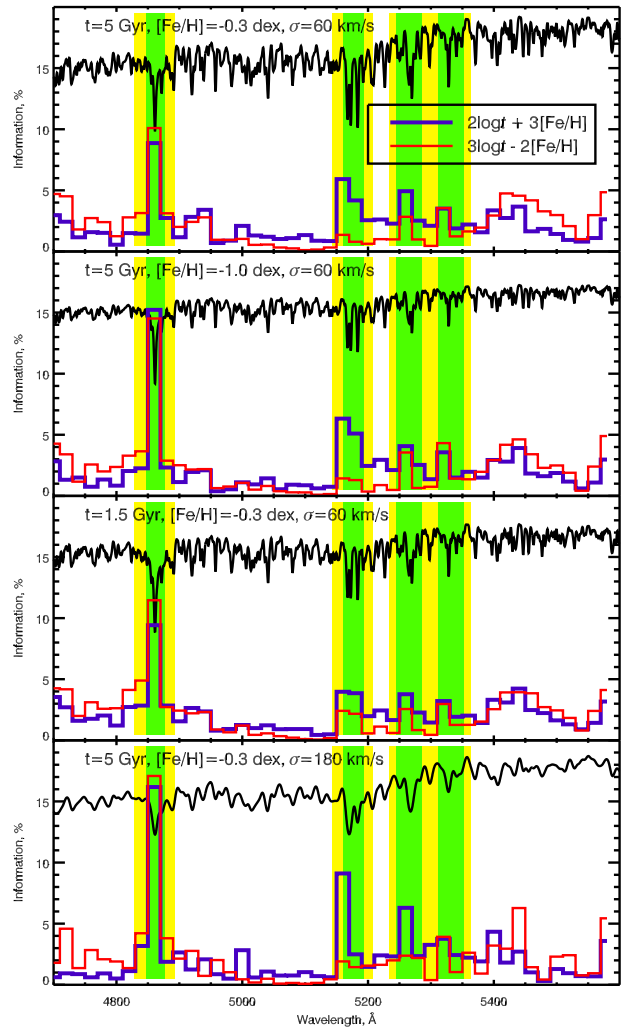}
\caption{Spectral distribution of the stellar population sensitive
information for four stellar populations in 20~\AA-wide bins. Corresponding
age, metallicity, and velocity dispersion are indicated in every panel. The
studied SSPs are shown in black; blue and red histograms display the
relative importance of spectral bins (in per cent) for the determination of
$\eta$ and $\theta$ stellar population properties respectively. Green bars
denote the bands defining the following Lick indices: H$\beta$, Mg$b$,
Fe$_{5270}$, and Fe$_{5335}$ with yellow side-bars corresponding to the
corresponding pseudo-continuum definition regions. \label{figspecinfo}}
\end{figure}

We have co-added the information on $\eta$-, and $\theta$-sensitivity in
20~\AA\ bins in the wavelength range between 4700 and 5600~\AA\ and
normalised it by the total value of non-reduced $\chi^2$ thus obtaining the
relative importance of every 20~\AA-wide bin to the determination of the
stellar population parameters. Four examples with different stellar
populations and velocity dispersions are provided in Fig~\ref{figspecinfo}.
One should not directly compare blue and red curves, because they represent
the normalised quantities: the absolute values of $S(\lambda, \eta)$ 
as defined by Eq.~\ref{senseq} are several times higher than $S(\lambda, \theta)$
explaining the elongated shapes of the 1-$\sigma$ uncertainty ellipses
in the age--metallicity space.

As expected, the most valuable information is contained in prominent
absorption-line features such as H$\beta$ and Mg$b$ having corresponding
Lick indices defined. However, it is noticeable that $\eta$ and
$\theta$-sensitive information is present everywhere in the spectrum
although in different amounts. Going to low metallicities or young ages,
thus, reducing the absorption-line features apart from H$\beta$, causes
concentration of information in the most prominent lines. Increasing the
intrinsic velocity dispersion also does so by smearing out faint spectral
features, therefore for high-resolution spectra of low velocity dispersion
objects it is possible to take the full advantage of techniques
such as {\sc NBursts} over line-strength indices.

To be stressed that in none of the presented examples the 20~\AA-wide bin
around H$\beta$ contains even 20~per cent of the $\theta$-sensitive
information. This explains why it is possible to precisely constrain the
stellar population age using full spectral fitting techniques even if
H$\beta$ (or other Balmer lines) are not available
\citep{CCB08,Chilingarian+08}.

To measure the metallicity gradients we fit the co-added spectra in the two
regions along the radius adding the data on both sides of the galaxy centre,
individually selected for every object. With our data it is not possible to
define the regions automatically based only on the values of $r_{e}$,
because the effective surface brightness values as well as exposure times
differ quite significantly from object to object, leading to very different
signal-to-noise ratios at a given distance from the centre in terms of
$r_{e}$. Therefore, we based the definitions of the radial zones based on
the observed radius of a nucleus $r_{nuc}$ defined as a half-size of a
pronounced bump in the metallicity profile, if observed or 1.5~arcsec
otherwise, resulting in a range of values from 1.5 to 4~arcsec (for
VCC~478). The inner radius of the inner region was chosen to be $r_{ii} =
1.5 r_{nuc}$ to prevent the contamination of the spectra by the light of a
galactic nucleus, but not smaller than 3.5~arcsec. The outer radius of the
inner region and the parameters of the outer region were defined as follows:
$r_{io} = (2.5 / 1.5) r_{ii}; r_{oi} = (3.5 / 1.5) r_{ii}; r_{oo} = (5.5 /
1.5) r_{ii}$. At the end, inner and outer regions turned to be about 1/4 to
1/2 and 1/2 to 1 in the units of effective radii. The values of the
metallicity gradients were computed as $\Delta[\mbox{Fe/H}]/\Delta r = 2
([\mbox{Fe/H}]_{o} - [\mbox{Fe/H}]_{i}) / (r_{oi} + r_{oo} - (r_{ii} +
r_{io}))$, where $[\mbox{Fe/H}]_{i}$ and $[\mbox{Fe/H}]_{o}$ are the values
of metallicity derived from the fitting of the co-added spectra in the inner
and outer regions along the radius. Then, the values were converted into
$\Delta \mbox{Fe/H}$ per $r_{e}$ and per kpc.

We computed the radial variations of the $B$-band stellar mass-to-light
ratios using the {\sc pegase.2} \citep{FR97} code for the Salpeter IMF. It
is important to mention that according the the predictions of {\sc
pegase.2}, in addition to the stellar content, another 30 to
40~per~cent of the mass is supposed to be contained in the form of ISM. The
cluster dwarf elliptical galaxies are supposed to efficiently lose their ISM
in quite short time by means of gravitational harassment (see e.g.
\citealp{Moore+96}) or ram-pressure stripping (see e.g. \citealp{AMB99}).
Therefore, to estimate the mass-to-light ratios of the stellar populations
we do not take into account this gas mass. The two other remaining
components of the total SSP mass, white dwarfs and neutron stars / black
holes are taken into account.

The parameters referred  throughout the paper as ``periphery'' or ``main
galactic body'' were obtained from the co-added spectra along the slit on
both sides of the photometric nuclei of galaxies at radii from $r_{ii}$
to $r_{oo}$.

For the objects having spectra with sufficient signal-to-noise ratios in the
blue spectral range we also measured the Lick indices. We compared the
measurements of Mg$b$ and the combined iron index $\langle \mbox{Fe} \rangle
= 0.72 \mbox{Fe}_{5270} + 0.28 \mbox{Fe}_{5335}$ \citep{TMB03} with the SSP
models for different $\alpha$-enhancement presented in the same study. These
measurements were made in the nuclear and peripheral parts of the galaxies,
as well as for all the SDSS DR6 spectra. To measure the Lick indices we have
transformed the data into the Lick system by convolving the spectra with the
quadratic difference of the Lick spectral resolution and the effective
spectral resolution of the data (instrumental resolution and the intrinsic
broadening due to velocity dispersion of the galaxies), as was proposed by
\cite{Kuntschner04} and applied to the SDSS data, e.g. in
\cite{Chilingarian+08}. \footnote{Since we did not have sufficient number of
observations of Lick standard stars, no systemic offsets between our data
and the original Lick/IDS response function (see e.g. Appendix~A of
\citealp{Michielsen+08}) could be calculated and applied to our measurements of
the Lick indices.} We derived the values of the $\alpha$/Fe abundance ratios
by inverting the grids of models presented in \cite{TMB03} using the age
values obtained from the full spectral fitting procedure described above.

\section{Results}

In Table~\ref{tabres1} we present the kinematical and stellar population
properties of the galaxies in our sample. We provide the global
photometrical parameters, absolute magnitudes $M_B$ corrected for the
Galactic extinction and half-light radii $r_e$ from the literature. The
conversion from the photometric systems in the original bibliographic
sources to the $B$ band was done using the transformations of \cite{FSI95}.
We compiled the photometric measurements from the following studies (see
Table~\ref{tabsample}): the ACS Virgo Cluster Survey \citep{Ferrarese+06},
the Keck dE project \citep{GGvdM02,GGvdM03}, HST WFPC2 surface photometry
\citep{Stiavelli+01}, Palomar dE project \citep{vZSH04,vZBS04}, SDSS DR6
\citep{SDSS_DR6}. From the SDSS DR6 catalogue we have adopted the values of
the $g'$-band 50-percentile Petrosian radius as $r_e$. We have used the
distance modulus of the Virgo cluster $-31.09$~mag, as in
\cite{Ferrarese+06}, and respectively corrected the absolute magnitudes from
other studies.

The following quantities are provided for the nuclear regions and main
bodies of the galaxies: velocity dispersion, luminosity-weighted age,
metallicity, stellar mass-to-light ratios assuming the Salpeter IMF, [Mg/Fe]
abundance ratio. We also provide metallicity gradients per kpc and $r_e$,
computed as described above.

Since the stellar population properties estimated by the full spectral
fitting are a subject to the age--metallicity degeneracy, we provide the
uncertainties in the rotated coordinate system (see Eq.~\ref{coordeq}). The
two pairs of columns, $\Delta \eta$ and $\Delta \theta$ provide a
possibility to recover the approximate dimensions of the inclined error
ellipses in the age--metallicity plot for the nuclear and peripheral parts
of the galaxies.

\begin{table*}
\caption{Kinematical and stellar population properties of dwarf early-type
galaxies: circumnuclear regions. The columns are as follows:
object name, absolute magnitude (extinction corrected), effective radius,
adopted nuclear radius, central velocity dispersion, age, metallicity,
2 columns representing uncertainties of stellar population parameters
(see text), central $[$Mg/Fe$]$ abundance ratio, $B$-band mass-to-light
ratio, 2 columns providing ages and metallicities from the fitting of 
SDSS DR6 spectra.\label{tabres1}}
\begin{tabular}{lcccccccccccc}
 VCC & $M_B$ & $r_{e}$ & $r_{n}$ & $\sigma_{0}$ & $t$ & [Fe/H] &
$\Delta \eta$ & $\Delta \theta$ & [Mg/Fe] & $(M/L)_{B}$ & $t_{\mbox{SDSS}}$ & [Fe/H]$_{\mbox{SDSS}}$\\
 & mag & arcsec & arcsec & km~s~$^{-1}$ & Gyr & dex & dex & dex & dex & 
$(M/L)_{\odot}$ & Gyr & dex \\
\hline
\hline
0178 & -15.43 &   8.7 &   2.0 &  50$\pm$2 & 10.6 & -0.73 &  0.08 &  0.03 &  0.14$\pm$0.05 &  4.8 & 8.5$\pm$2.0 &  -0.81$\pm$0.04 \\
0389$^a$ & -16.28 &  10.9 &   2.5 &  23$\pm$3 &  3.9 & -0.38 &  0.09 &  0.07 &$\ldots\pm\ldots$ &  2.8 & 3.6$\pm$0.4 &  -0.27$\pm$0.03 \\
0389$^a$ & -16.28 &  10.9 &   2.5 &  36$\pm$2 &  4.9 & -0.30 &  0.10 &  0.04 &$\ldots\pm\ldots$ &  3.6 & 3.6$\pm$0.4 &  -0.27$\pm$0.03 \\
0437 & -16.85 &  25.1 &   4.0 &  47$\pm$1 &  4.5 & -0.36 &  0.04 &  0.02 &  0.06$\pm$0.05 &  3.2 & $\ldots$ & $\ldots$ \\
0490$^a$ & -16.14 &  17.4 &  17.4 &  33$\pm$5 &  5.1 & -0.23 &  0.23 &  0.09 &$\ldots\pm\ldots$ &  3.9 & 2.4$\pm$0.3 &  -0.15$\pm$0.03 \\
0490$^{b}$ & -16.14 &  17.4 &  1.5 &  34$\pm$8 &  3.3 & -0.35 &  0.13 &  0.04 &$\ldots\pm\ldots$&  2.4 & 2.4$\pm$0.3 &  -0.15$\pm$0.03 \\
0543 & -16.55 &  17.1 &   2.0 &  49$\pm$1 &  5.4 & -0.49 &  0.06 &  0.01 &  0.09$\pm$0.06 &  3.2 & 4.9$\pm$0.7 &  -0.30$\pm$0.03 \\
0543$^a$ & -16.55 &  17.9 &   2.0 &  48$\pm$3 &  5.7 & -0.41 &  0.14 &  0.04 &$\ldots\pm\ldots$ &  3.6 & 4.9$\pm$0.7 &  -0.30$\pm$0.03 \\
0545$^a$ & -15.10 &   9.5 &   9.5 &  31$\pm$7 &  5.1 & -0.47 &  0.27 &  0.07 &$\ldots\pm\ldots$&  3.2 &  9.3$\pm$2.9 &  -0.79$\pm$0.05 \\
0634$^a$ & -16.31 &  16.6 &   2.5 &  26$\pm$2 &  4.0 & -0.30 &  0.13 &  0.07 &$\ldots\pm\ldots$&  3.1 &  2.5$\pm$0.3 &  -0.17$\pm$0.03 \\
0856$^a$ & -16.42 &  14.7 &   2.5 &  31$\pm$1 &  5.0 & -0.43 &  0.10 &  0.03 &  0.06$\pm$0.12 &  3.2 &  4.1$\pm$0.6 &  -0.39$\pm$0.03 \\
0917 & -15.48 &   7.8 &   2.0 &  41$\pm$1 &  4.4 & -0.62 &  0.05 &  0.02 &  0.03$\pm$0.04 &  2.6 &  4.9$\pm$0.8 &  -0.48$\pm$0.02 \\
0965 & -15.44 &  13.8 &   1.5 &  34$\pm$4 &  2.2 & -0.56 &  0.12 &  0.04 & -0.14$\pm$0.07 &  1.4 &  3.0$\pm$1.2 &  -0.66$\pm$0.10 \\
0990 & -16.31 &   9.7 &   2.0 &  53$\pm$1 &  4.0 & -0.40 &  0.03 &  0.02 &  0.06$\pm$0.03 &  2.8 &  4.1$\pm$0.5 &  -0.32$\pm$0.03 \\
1010$^a$ & -16.79 &  22.9 &   2.5 &  43$\pm$2 &  7.3 & -0.33 &  0.10 &  0.04 &  0.03$\pm$0.09 &  4.8 & $\ldots$ & $\ldots$ \\
1036 & -17.03 &  14.6 &   1.5 &  53$\pm$1 &  3.5 & -0.15 &  0.02 &  0.01 & -0.05$\pm$0.03 &  3.0 &  3.0$\pm$0.2 &  -0.02$\pm$0.02 \\
1036$^a$ & -17.03 &  14.6 &   1.5 &  41$\pm$2 &  2.6 & -0.02 &  0.09 &  0.04 &$\ldots\pm\ldots$&  2.4 &  3.0$\pm$0.2 &  -0.02$\pm$0.02 \\
1075 & -15.71 &  14.6 &   1.5 &  53$\pm$7 &  5.7 & -0.74 &  0.22 &  0.06 &$\ldots\pm\ldots$&  2.8 &  7.9$\pm$3.6 &  -0.77$\pm$0.07 \\
1087$^a$ & -16.80 &  18.9 &   2.5 &  63$\pm$4 &  5.3 & -0.27 &  0.13 &  0.04 &$\ldots\pm\ldots$&  3.9 &  5.2$\pm$0.8 &  -0.22$\pm$0.02 \\
1122 & -16.03 &  11.4 &   2.5 &  40$\pm$1 &  2.4 & -0.37 &  0.04 &  0.01 &  0.05$\pm$0.05 &  1.7 &  2.6$\pm$0.4 &  -0.28$\pm$0.02 \\
1122$^a$ & -16.03 &  11.4 &   2.5 &  33$\pm$1 &  2.4 & -0.30 &  0.09 &  0.03 &$\ldots\pm\ldots$&  1.9 &  2.6$\pm$0.4 &  -0.28$\pm$0.02 \\
1183$^a$ & -16.66 &  10.5 &   2.5 &  56$\pm$2 &  4.7 & -0.28 &  0.08 &  0.03 &$\ldots\pm\ldots$&  3.6 & $\ldots$ & $\ldots$ \\
1250$^a$ & -18.27 &  16.2 &   2.5 &  79$\pm$4 &  2.2 & -0.07 &  0.11 &  0.04 & -0.02$\pm$0.15 &  2.0 &  2.8$\pm$0.1 &  -0.05$\pm$0.01 \\
1261 & -17.38 &  21.1 &   2.5 &  54$\pm$1 &  2.5 & -0.27 &  0.02 &  0.01 & -0.06$\pm$0.03 &  2.0 &  2.6$\pm$0.3 &  -0.25$\pm$0.02 \\
1308 & -15.41 &   9.1 &   2.0 &  47$\pm$1 &  3.6 & -0.46 &  0.04 &  0.03 & -0.00$\pm$0.07 &  2.4 & $\ldots$ & $\ldots$ \\
1407$^a$ & -15.64 &   8.9 &   8.9 &  36$\pm$5 &  5.0 & -0.51 &  0.31 &  0.07 &$\ldots\pm\ldots$&  3.1 &  5.8$\pm$1.5 &  -0.54$\pm$0.03 \\
1422$^{b}$ & -17.07 &  19.3 &  1.5 &  44$\pm$7 &  7.1 & -0.48 &  0.09 &  0.04 &$\ldots\pm\ldots$&  4.1 &  5.5$\pm$0.8 &  -0.30$\pm$0.02 \\
1491$^a$ & -15.44 &  10.2 &   2.5 &  47$\pm$4 &  5.6 & -0.44 &  0.18 &  0.05 &$\ldots\pm\ldots$&  3.5 &  5.4$\pm$1.2 &  -0.33$\pm$0.03 \\
1514 & -15.58 &  13.7 &   2.0 &  42$\pm$3 &  4.5 & -0.77 &  0.09 &  0.04 & -0.04$\pm$0.05 &  2.3 & $\ldots$ & $\ldots$ \\
1545$^{b}$ & -15.90 &  11.2 &  1.5 &  59$\pm$5 &  4.3 & -0.05 &  0.07 & 0.03 &$\ldots\pm\ldots$&  3.9 &  5.5$\pm$0.6 &  -0.16$\pm$0.02 \\
1743 & -15.20 &  10.4 &  10.4 &  44$\pm$8 & 14.0 & -1.01 &  0.13 &  0.12 &$\ldots\pm\ldots$&  5.0 &  2.9$\pm$1.0 &  -0.46$\pm$0.08 \\
1857 & -15.85 &  20.8 &  20.8 &  46$\pm$16 &  2.7 & -0.66 &  0.42 &  0.15 &$\ldots\pm\ldots$&  1.5 & $\ldots$ & $\ldots$ \\
1871$^{b}$ & -16.54 &   6.9 &  1.5 &  68$\pm$3 &  4.9 & +0.03 &  0.04 & 0.02 & -0.02$\pm$0.03&  4.7 &  5.7$\pm$0.4 & +0.11$\pm$0.01 \\
2019 & -16.25 &  15.4 &   1.5 &  44$\pm$1 &  2.3 & -0.28 &  0.05 &  0.01 & -0.03$\pm$0.07 &  1.8 & $\ldots$ & $\ldots$ \\
2048$^a$ & -16.85 &  12.6 &   2.5 &  33$\pm$2 &  3.9 & -0.44 &  0.06 &  0.05 & -0.01$\pm$0.15 &  2.7 & $\ldots$ & $\ldots$ \\
2050 & -15.55 &  10.8 &   2.5 &  32$\pm$2 &  2.6 & -0.54 &  0.06 &  0.02 & -0.17$\pm$0.09 &  1.7  &  2.5$\pm$0.6 &  -0.28$\pm$0.04\\
\hline
\hline
\end{tabular}

\flushleft

\footnotesize{$^a$ Data from the HyperLEDA FITS archive (OHP CARELEC).}

\footnotesize{$^b$ IFU spectroscopic data (MPFS).}
\end{table*}

\begin{table*}
\caption{Kinematical and stellar population properties of dwarf early-type
galaxies: external regions. The columns are as follows: object name,
velocity dispersion in the periphery, age, metallicity, 2 columns
representing uncertainties of stellar population parameters
(see text), $[$Mg/Fe$]$ abundance ratio, $B$-band mass-to-light ratio,
metallicity gradients per r$_e$ and per $kpc$ where positive values denote 
the decrease of [Fe/H] outwards.
\label{tabres2}}
\begin{tabular}{lccccccccc}
 VCC & $\sigma_{p}$ & $t$ & [Fe/H] &
$\Delta \eta$ & $\Delta \theta$ & [Mg/Fe] & $(M/L)_{B}$ &
$\Delta[\mbox{Fe/H}]/\Delta r$ & $\Delta[\mbox{Fe/H}]/\Delta r$ \\
 & km~s~$^{-1}$ & Gyr & dex & dex & dex & dex & $(M/L)_{\odot}$ &
dex~$r_{e}^{-1}$ & dex~kpc$^{-1}$\\
\hline
\hline
0178 &  39$\pm$3 & 11.5 & -0.83 &  0.09 &  0.05 &  0.07$\pm$0.05 &  4.8 &  0.48$\pm$0.20 &  0.68$\pm$0.29 \\
0389$^a$ &  17$\pm$6 &  3.7 & -0.35 &  0.15 &  0.13 & $\ldots\pm\ldots$&  2.7 &$\ldots\pm\ldots$ &$\ldots\pm\ldots$ \\
0389$^a$ &  23$\pm$4 &  3.2 & -0.44 &  0.17 &  0.15 & $\ldots\pm\ldots$&  2.1 &$\ldots\pm\ldots$ &$\ldots\pm\ldots$ \\
0437 &  44$\pm$3 & 17.9 & -1.04 &  0.03 &  0.05 &  0.20$\pm$0.07 &  5.8 &  0.91$\pm$0.11 &  0.45$\pm$0.05 \\
0490$^a$ &  $\ldots\pm\ldots$ &  $\ldots$ & $\ldots$ &  $\ldots$ & $\ldots$ & $\ldots\pm\ldots$ & $\ldots$ & $\ldots\pm\ldots$ & $\ldots\pm\ldots$ \\
0490$^{b}$ & 29$\pm$7 &  12.8 & -0.79 &  0.23 & 0.12 & $\ldots\pm\ldots$ & 5.4 & $\ldots\pm\ldots$ & $\ldots\pm\ldots$ \\
0543 &  44$\pm$2 &  7.3 & -0.69 &  0.07 &  0.03 &  0.16$\pm$0.06 &  3.6 &  0.37$\pm$0.17 &  0.27$\pm$0.13 \\
0543$^a$ &  48$\pm$4 &  6.4 & -0.63 &  0.17 &  0.06 &  $\ldots\pm\ldots$&  3.3 &$\ldots\pm\ldots$&$\ldots\pm\ldots$ \\
0545$^a$ &  $\ldots\pm\ldots$ &  $\ldots$ & $\ldots$ &  $\ldots$ & $\ldots$ & $\ldots\pm\ldots$ & $\ldots$ & $\ldots\pm\ldots$ & $\ldots\pm\ldots$ \\
0634$^a$ &  31$\pm$4 &  2.1 & -0.11 &  0.12 &  0.04 & $\ldots\pm\ldots$&  1.9 &$\ldots\pm\ldots$&$\ldots\pm\ldots$\\
0856$^a$ &  31$\pm$2 &  5.9 & -0.45 &  0.14 &  0.04 & $\ldots\pm\ldots$&  3.6 &$\ldots\pm\ldots$&$\ldots\pm\ldots$ \\
0917 &  42$\pm$3 & 10.9 & -1.00 &  0.10 &  0.05 & -0.01$\pm$0.06 &  4.2 &  0.21$\pm$0.19 &  0.33$\pm$0.31 \\
0965 &  12$\pm$15 &  6.5 & -1.01 &  0.17 &  0.07 &$\ldots\pm\ldots$&  2.7 &  0.55$\pm$0.29 &  0.50$\pm$0.26 \\
0990 &  51$\pm$1 &  5.1 & -0.59 &  0.06 &  0.01 &  0.16$\pm$0.04 &  2.9 &  0.20$\pm$0.06 &  0.25$\pm$0.08 \\
1010$^a$ &  49$\pm$2 &  7.7 & -0.40 &  0.12 &  0.04 & -0.06$\pm$0.11 &  4.8 &$\ldots\pm\ldots$ &$\ldots\pm\ldots$\\
1036 &  56$\pm$1 &  4.9 & -0.33 &  0.03 &  0.01 &  0.17$\pm$0.03 &  3.5 &  0.48$\pm$0.06 &  0.41$\pm$0.05 \\
1036$^a$ &  43$\pm$3 &  2.9 & -0.13 &  0.11 &  0.04 & $\ldots\pm\ldots$ &  2.4 &$\ldots\pm\ldots$ &$\ldots\pm\ldots$ \\
1075 &  41$\pm$7 &  5.0 & -0.52 &  0.22 &  0.05 &$\ldots\pm\ldots$&  3.1 & -0.09$\pm$0.23 & -0.08$\pm$0.19 \\
1087$^a$ &  65$\pm$5 & 13.8 & -0.60 &  0.21 &  0.15 & $\ldots\pm\ldots$ &  6.5 &$\ldots\pm\ldots$ &$\ldots\pm\ldots$ \\
1122 &  39$\pm$2 &  6.7 & -0.82 &  0.07 &  0.03 &  0.09$\pm$0.06 &  3.1 &  0.38$\pm$0.10 &  0.42$\pm$0.11 \\
1122$^a$ &  35$\pm$3 &  3.8 & -0.55 &  0.08 &  0.08 & $\ldots\pm\ldots$ &  2.3 &$\ldots\pm\ldots$ &$\ldots\pm\ldots$ \\
1183$^a$ &  65$\pm$5 &  8.3 & -0.52 &  0.17 &  0.06 & $\ldots\pm\ldots$&  4.6 & $\ldots\pm\ldots$& $\ldots\pm\ldots$\\
1250$^a$ &  80$\pm$3 &  2.3 & -0.05 &  0.08 &  0.03 &  0.11$\pm$0.11 &  2.1 &  $\ldots\pm\ldots$ &  $\ldots\pm\ldots$\\
1261 &  57$\pm$1 &  4.9 & -0.61 &  0.05 &  0.01 &  0.24$\pm$0.04 &  2.8 &  0.40$\pm$0.10 &  0.24$\pm$0.06 \\
1308 &  45$\pm$3 &  5.9 & -0.75 &  0.09 &  0.02 &  0.02$\pm$0.06 &  2.9 &  0.38$\pm$0.10 &  0.52$\pm$0.13 \\
1407$^a$ &  $\ldots\pm\ldots$ &  $\ldots$ & $\ldots$ &  $\ldots$ & $\ldots$ & $\ldots\pm\ldots$ & $\ldots$ & $\ldots\pm\ldots$ & $\ldots\pm\ldots$ \\
1422$^{b}$ & 50$\pm$3 &   4.8 & -0.40 &  0.03 & 0.01 & $\ldots\pm\ldots$ & 3.3 & $\ldots\pm\ldots$ & $\ldots\pm\ldots$ \\
1491$^a$ &  37$\pm$8 &  7.4 & -0.57 &  0.31 &  0.14 & $\ldots\pm\ldots$ &  4.0 &$\ldots\pm\ldots$ & $\ldots\pm\ldots$\\
1514 &  48$\pm$3 &  3.0 & -0.64 &  0.09 &  0.07 &  0.11$\pm$0.06 &  1.7 &$\ldots\pm\ldots$&$\ldots\pm\ldots$\\
1545$^{b}$ & 64$\pm$4 &   8.3 & -0.40 &  0.06 & 0.03 & $\ldots\pm\ldots$ & 5.1 & $\ldots\pm\ldots$ & $\ldots\pm\ldots$ \\
1743 &  $\ldots\pm\ldots$ &  $\ldots$ & $\ldots$ &  $\ldots$ & $\ldots$ & $\ldots\pm\ldots$ & $\ldots$ & $\ldots\pm\ldots$ & $\ldots\pm\ldots$ \\
1857 &  $\ldots\pm\ldots$ &  $\ldots$ & $\ldots$ &  $\ldots$ & $\ldots$ & $\ldots\pm\ldots$ & $\ldots$ & $\ldots\pm\ldots$ & $\ldots\pm\ldots$ \\
1871$^{b}$ & 65$\pm$4 &  5.0 & -0.14 &  0.06 & 0.03 & 0.01$\pm$0.06 & 4.2 & $\ldots\pm\ldots$ & $\ldots\pm\ldots$ \\
2019 &  42$\pm$3 &  3.5 & -0.54 &  0.05 &  0.05 & -0.04$\pm$0.10 &  2.2 & -0.19$\pm$0.16 & -0.15$\pm$0.13 \\
2048$^a$ &  39$\pm$3 &  2.9 & -0.45 &  0.10 &  0.04 & -0.09$\pm$0.13 &  1.9 &$\ldots\pm\ldots$ & $\ldots\pm\ldots$\\
2050 &  21$\pm$5 &  4.5 & -0.76 &  0.11 &  0.04 &  0.03$\pm$0.09 &  2.3 &  0.20$\pm$0.16 &  0.23$\pm$0.19 \\
\hline
\hline
\end{tabular}
\flushleft

\footnotesize{$^a$ Data from the HyperLEDA FITS archive (OHP CARELEC).}

\footnotesize{$^b$ IFU spectroscopic data (MPFS).}
\end{table*}

\subsection{Comparison of the Datasets}

\begin{figure}
\includegraphics[width=\hsize]{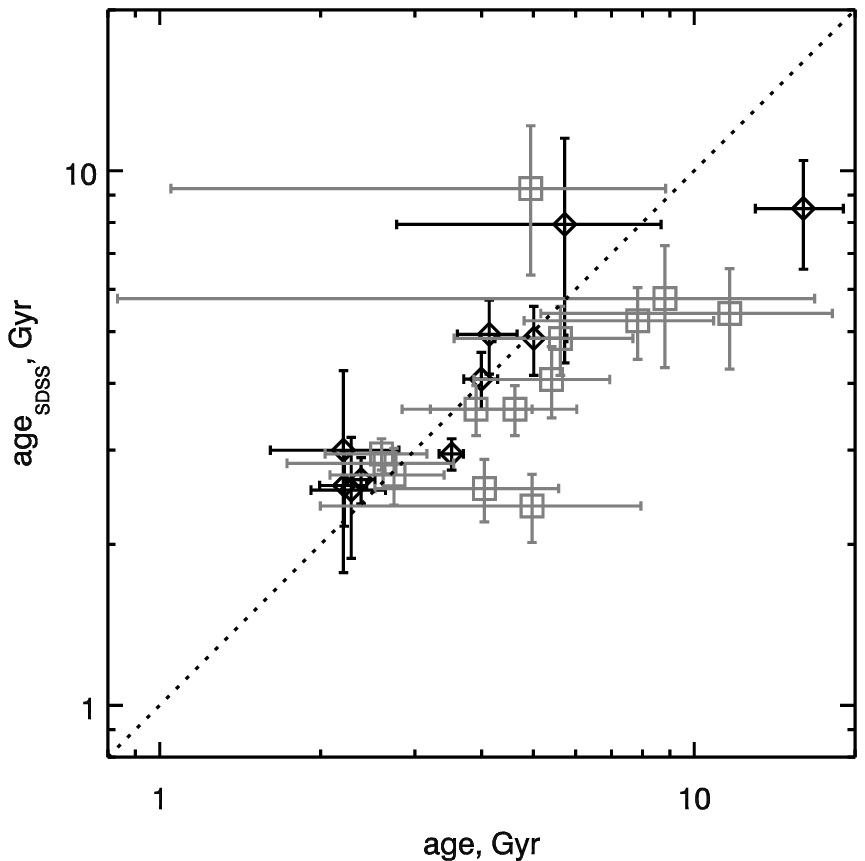} \\
\includegraphics[width=\hsize]{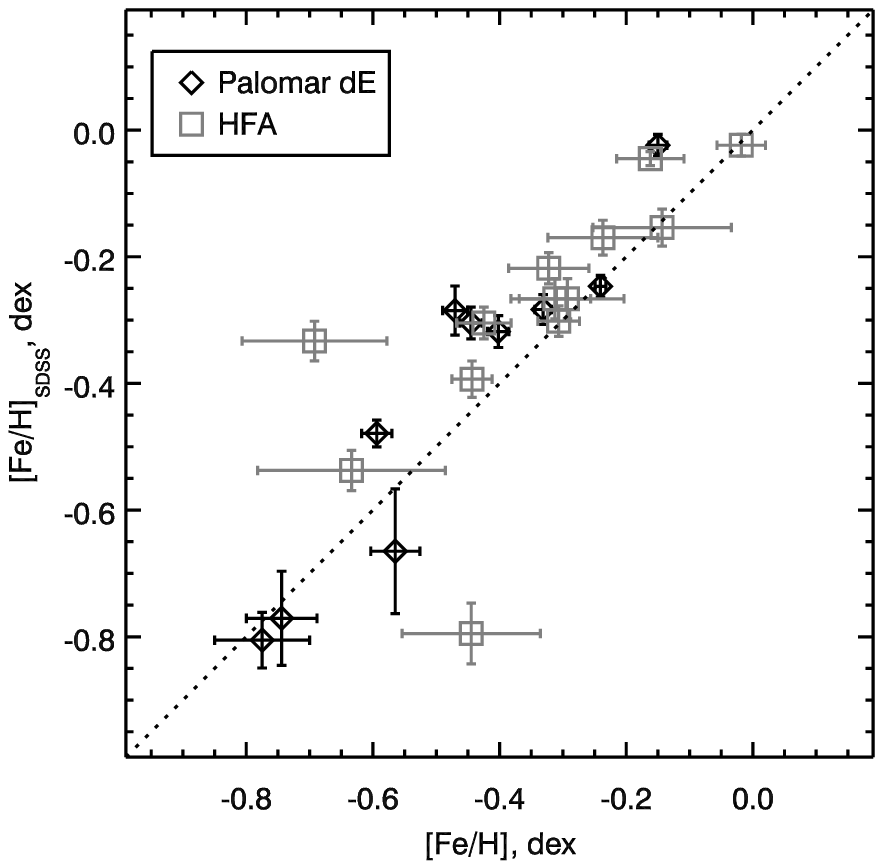}
\caption{Comparison of age (top panel) and metallicity (bottom panel)
measurements in the central parts of the galaxies made on the SDSS DR6
spectra and on the long-slit data. Different symbols correspond
to the data from the Palomar dE project and HyperLeda FITS 
Archive.
\label{figSDSScomp}
}
\end{figure}

For the long-slit and IFU datasets we have fit the co-added spectra of the
central regions of the galaxies in the 3~arcsec-wide aperture in order to
compare the derived stellar population properties directly with those
obtained from the fitting of the SDSS spectra. In Fig~\ref{figSDSScomp} we
compare these measurements. The agreement is very good showing no systematic
offsets between the different datasets, which suggests that the data were
reduced properly. Since there is no systematic offset in the metallicity
estimates, we conclude that the subtraction of additive background
components, such as diffuse light in the spectrograph or the sky background,
was precise. This is an important conclusion, that SDSS spectra allow us to
study the stellar population properties of dE/dS0 nuclei. Although the
spectra are contaminated by the light of main galactic discs/spheroids, the
results obtained from the full spectral fitting of SDSS spectra are fairly
consistent with those obtained from the analysis of other ground-based data,
giving possibility to study much broader samples of dwarf early-type galaxies
using only SDSS data.

For the three objects, VCC~543, VCC~1036, and VCC~1122 having long-slit data
in both, Palomar dE project and the HyperLeda FITS Archive, the agreement
between the derived kinematical and stellar population profiles is fairly
good (see details in the following subsection on individual galaxies).

We quantitatively compare our kinematical profiles with the results
published in \cite{SP02} which are available in the electronic form. In all
the cases both, the measurements and the derived uncertainties are
absolutely consistent between the two studies although they were obtained
using completely different data analysis techniques. We notice that in
order to derive the stellar population properties we need higher
signal-to-noise ratio than required only to constraint kinematics,
therefore we had to apply coarser binning to the data degrading the spatial
resolution to a higher degree than that presented in \cite{SP02}.

Since no kinematical profiles are provided in the computer-readable form in
the other publications presenting dE/dS0 kinematics, we were able to make
only qualitative comparison with the published kinematical profiles for some
objects.

\cite{vZSH04} present two kinematical profiles for every galaxy observed in
the course of the Palomar dE project, corresponding to the spectra obtained
in the blue and red arms of the Double Spectrograph. The red arm data, having
higher spectral resolution and sharper and deeper absorption lines of the
Ca~II triplet result in higher quality kinematical data, which is clearly
seen in Fig~3 of \cite{vZSH04}. It is also clear from the same figure that
the velocity dispersion values obtained from the blue arm spectra are
generally systematically above those derived from the red arm data, which is
probably caused by the imperfections of the data analysis in low-$\sigma$
regime, where measured velocity dispersions are similar to or lower than the
instrumental resolution. Our analysis of the blue arm spectra in terms of
kinematics turns to be much more precise than what is presented in
\cite{vZSH04}. In the central regions of the galaxies our velocity
dispersion profiles agree remarkably well with those derived from the higher
resolution red arm spectra by \cite{vZSH04}, whereas in the peripheral parts
we usually reach even higher quality of measurements evident from the lower
scatter of the values. The two exceptions are VCC~956 and VCC~2050, where
velocity dispersions are too low to be extracted from blue arm spectra.

The kinematical profiles of several galaxies included in our sample were
presented in the literature. The rotation profiles agree well for VCC~543
and VCC~1010 displayed in Fig~2 of \cite{Pedraz+02}, while in VCC~1122 we see
somewhat higher degree of rotation. The velocity dispersion values are
consistent for all three objects. Their data were obtained with the IDS
spectrograph at the 2.5 Isaac Newton Telescope and had $\sim$2.5\AA\ FWHM
spectral resolution in the blue spectral range, which is very similar to the
datasets we discuss in this paper.

Seven galaxies included in our sample, VCC~543, 856, 917, 1036, 1087, 1261,
and 1308 were observed with the Keck ESI spectrograph, and the kinematical
analysis was presented in \cite{GGvdM02,GGvdM03}. The data had considerably
higher spectral resolution than ours ($R \sim 10000$) but were localised in
the central regions of the galaxies (out to 10~arcsec from the centres) due
to a relatively short slit. Both, radial velocity and velocity dispersion
profiles for the seven galaxies generally agree, however we notice that the
specific details (e.g. $\sigma$-drops) look smeared in the ESI data compared
to the Palomar DS spectra, although one would expect the opposite, given
better atmosphere conditions at Mauna Kea. Presently, we do not have an
explanation to this effect, but our guess is that it may be connected to
the atmosphere dispersion which would be an order of a few arcsec on a very
wide wavelength range of ESI even at moderate air-masses and might smooth
the kinematical details if one fits the spectra in the entire wavelength
range of the instrument. There is no mentioning about the atmosphere
dispersion corrector in the ESI manual available online, neither any
information about how it was taken into account in \cite{GGvdM02,GGvdM03}.
Comparing the shapes of the velocity dispersion profiles for VCC~917, 1036,
and 1261 in Fig~5 in \cite{GGvdM02} and Fig~2 and Fig~3 in \cite{GGvdM03},
to the data presented in our study one would understand why Geha et al. did
not mention the kinematically-decoupled cores in these galaxies: they were
smoothed out.

No spatially-resolved information about the stellar population of the
galaxies in our sample is available in the literature. Qualitative
comparison of the integrated values with those presented in \cite{GGvdM03}
and \cite{vZBS04} based on the analysis of the line-strength indices results
in a good agreement. The statistical comparison of the dE/dS0 stellar
population properties with another samples will be given in one of the
next papers of the series.

\subsection{Notes on Individual Galaxies}

Here we briefly describe the results obtained for every galaxy. All cases
with spatially-resolved information available are illustrated with the
profiles or two-dimensional maps of kinematical and stellar population
properties, except VCC~1871, where the colour figures are available in
\cite{CPSA07}.

\begin{figure}
\includegraphics[width=\hsize]{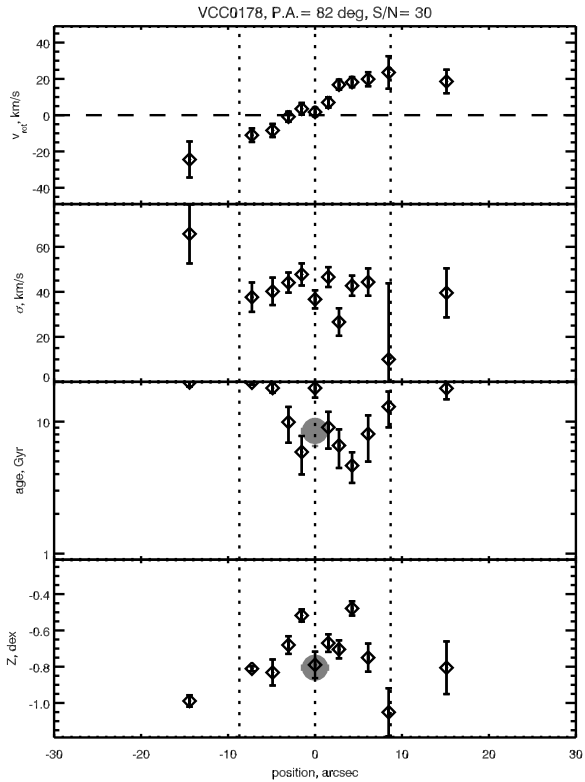}
\caption{Kinematics and stellar populations of VCC~178 (Palomar DS). Four panels (from
top to bottom) present: radial velocity, velocity dispersion, SSP-equivalent
age, and metallicity. Results from the fitting of a SDSS spectrum are shown
by filled gray circle. Positional angle of the slit and target
signal-to-noise ratio used to adaptively bin the data are indicated. Dotted
vertical lines indicate the photometric centre and $\pm r_e$.
\label{figvcc0178}}
\end{figure}

\noindent\emph{VCC~178 = IC~3081} (Fig~\ref{figvcc0178}). This flattened
nucleated dwarf galaxy exhibits significant rotation and quite flat velocity
dispersion profile inside 1~$r_{e}$. No embedded substructures were revealed
by \cite{LGB06}: VCC~178 is listed in Appendix~E there. There is a
significant metallicity gradient. Age distribution also demonstrates smooth
changes along the radius, being younger in the centre (7~Gyr) than in the
periphery (11.5~Gyr).

\begin{figure}
\includegraphics[width=\hsize]{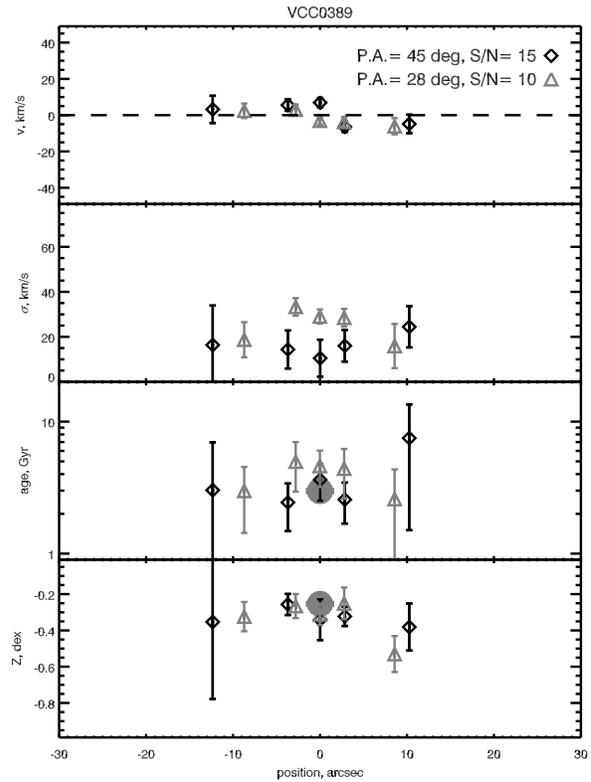}
\caption{Kinematics and stellar populations of VCC~389 (OHP CARELEC).
See Fig~\ref{figvcc0178} for details. Data from two datasets for slightly
different positional angles are shown with diamonds and triangles.
\label{figvcc0389}}
\end{figure}

\noindent\emph{VCC~389 = IC~781} (Fig~\ref{figvcc0389}). This nucleated dE
galaxy hosting a ``probable disc'' reported by \cite{LGB06} was observed
twice with the CARELEC spectrograph at slightly different positional angles.
Both datasets have quite low signal-to-noise ratios, however the derived
profiles of the stellar population parameters along the radius agree well.
We detected neither statistically significant difference of the stellar
population properties in the nucleus and in the main galactic body, nor
metallicity gradient.

\begin{figure}
\includegraphics[width=\hsize]{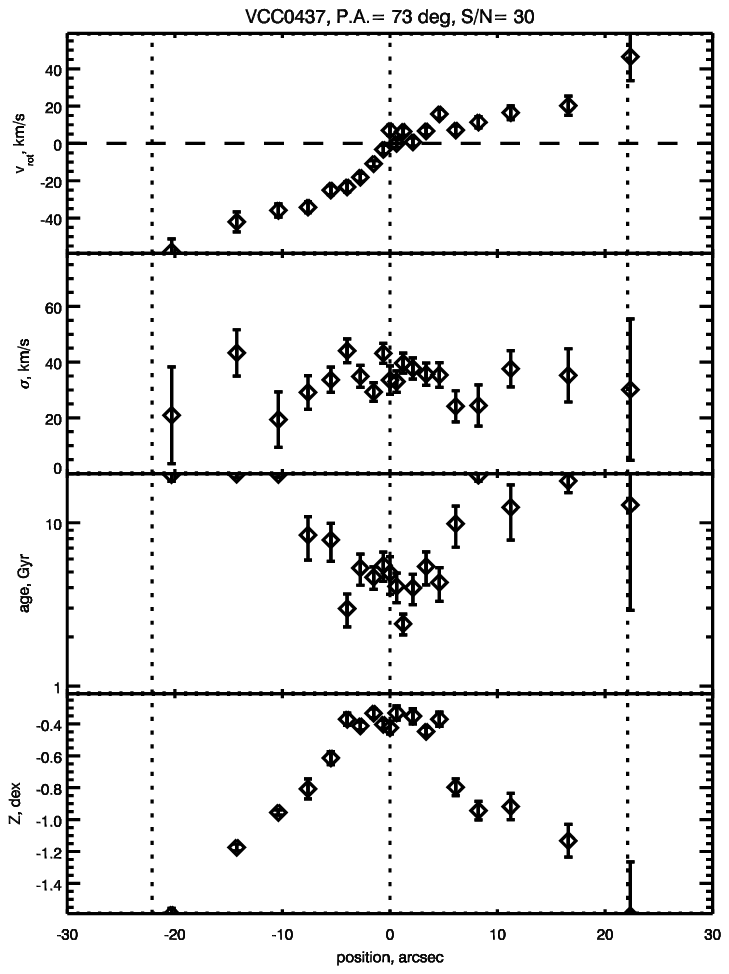}
\caption{Kinematics and stellar populations of VCC~437 (Palomar DS).
See Fig~\ref{figvcc0178} for details. \label{figvcc0437}}
\end{figure}

\noindent\emph{VCC~437 = UGC~7399A} (Fig~\ref{figvcc0437}). A very steep
metallicity gradient is observed beyond 4~arcsec from the centre.
Metallicity at 1~$r_e$ is almost 1~dex lower than at 4~arcsec.
Interestingly, at $\sim$3~arcsec the $g'-z'$ colour profile of the galaxy
presented in \cite{Ferrarese+06} changes its gradient, and the ellipticity
changes drastically. This suggests the presence of an embedded structure in
the central part, however, not clearly detected on the unsharp-masked SDSS
images, which can be explained if it has nearly face-on orientation. This
object is one of the cases, where the luminosity-weighted age also changes
along the radius: the population becomes older at larger radii, creating a
``conspiracy'' effect in the broad-band optical colours, making colour
profiles looking nearly flat.

\begin{figure}
\begin{tabular}{cc}
\includegraphics[width=0.5\hsize]{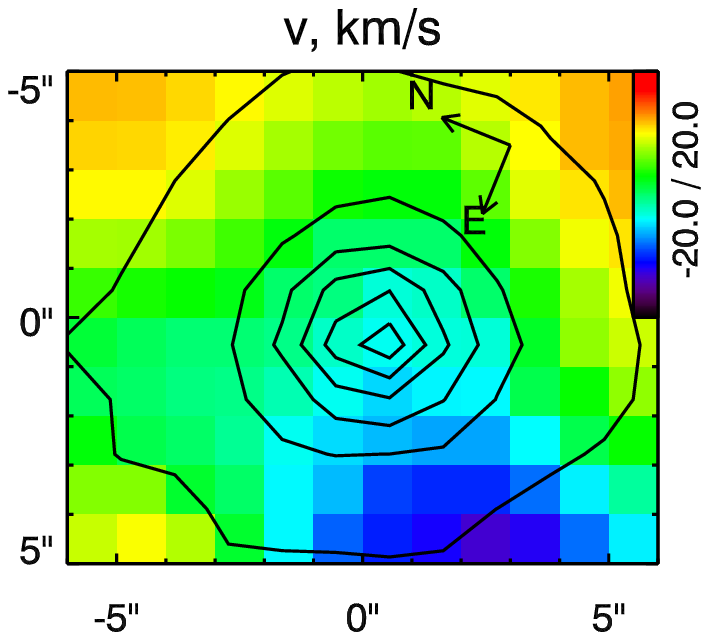} & 
\includegraphics[width=0.5\hsize]{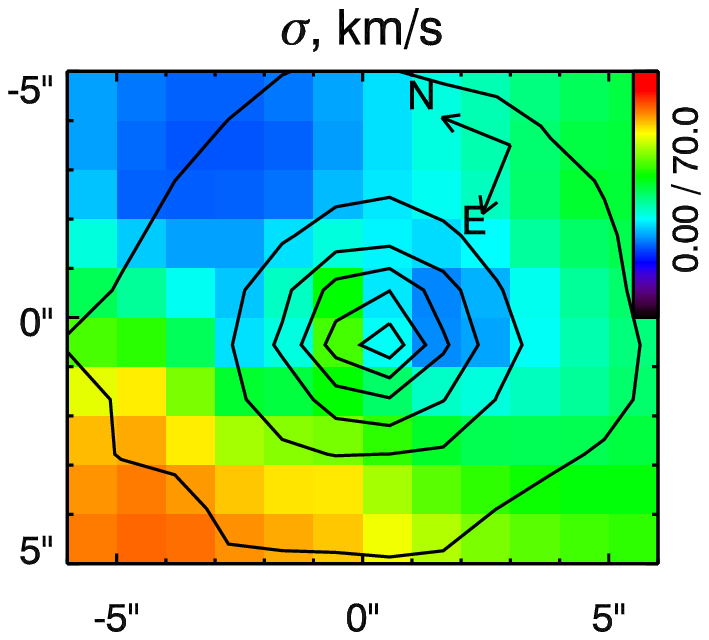} \\
\includegraphics[width=0.5\hsize]{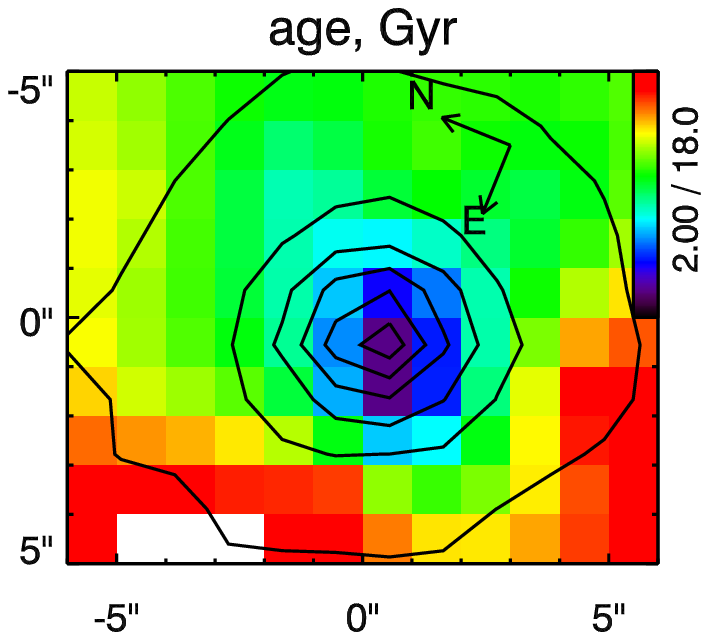} & 
\includegraphics[width=0.5\hsize]{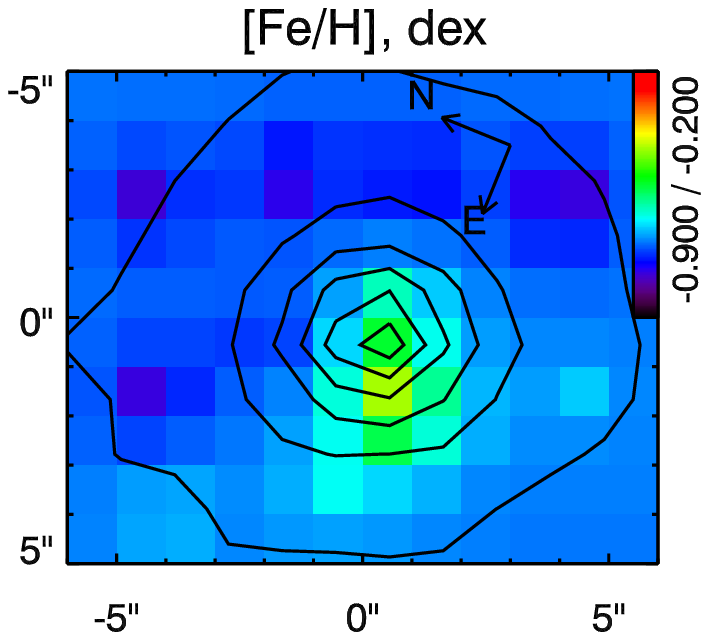} \\
\end{tabular}
\caption{Kinematics and stellar populations of VCC~490 (MPFS).
First row: radial velocity and velocity dispersion, second row: age and
metallicity. Adaptive binning with the target signal-to-noise ratio of 15
was applied.\label{figvcc0490}}
\end{figure}

\noindent\emph{VCC~490 = IC~783} (Fig~\ref{figvcc0490}). This nucleated dwarf
galaxy exhibits prominent spiral arms in the stellar disc
\citep{BBJ02,LGB06}. The HFA data available for this object have very low
signal-to-noise ratio outside the nuclear region, therefore the stellar
population properties cannot be derived in the periphery of the galaxy. Deep
IFU observations were carried out with the MPFS spectrograph in 2004 and
presented in \cite{Chilingarian06} and \cite{CPSA07}. The
luminosity-weighted age in the nuclear region of VCC~490 is as young as
3~Gyr, whereas the main galactic body is older than 10~Gyr. There is a hint of
a large-scale solid body rotation of the spiral disc \citep{SP02}. The
galaxy resides in the periphery of the Virgo cluster, close in projection
to the luminous spiral M~100. Their radial velocities are different by only
270~km~s$^{-1}$ which may suggest the belonging of VCC~490 to the M~100
group.

\begin{figure}
\includegraphics[width=\hsize]{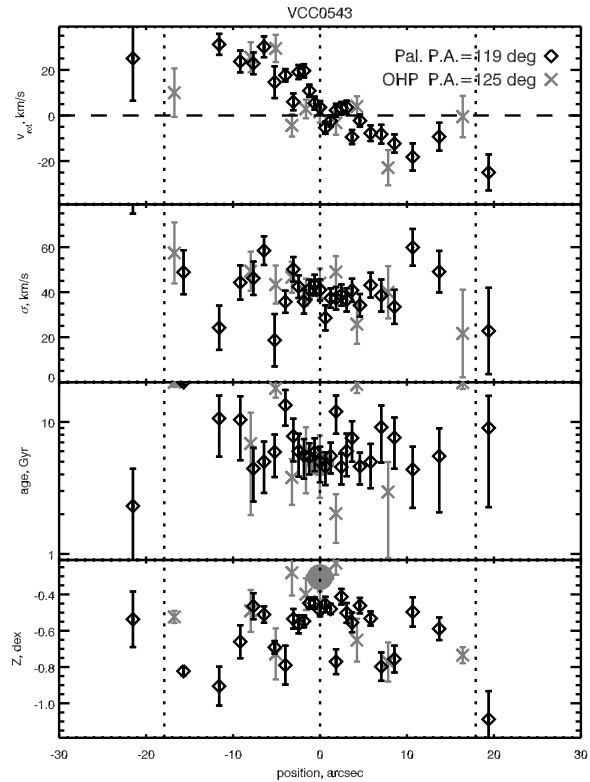}
\caption{Kinematics and stellar populations of VCC~543 (Palomar DS and OHP
CARELEC). Two curves on each panel correspond to the two different datasets.
The target signal-to-noise ratios were 20 and 10 for the Palomar DS and OHP
datasets respectively. See Fig~\ref{figvcc0178} for details. \label{figvcc0543}}
\end{figure}

\noindent\emph{VCC~543 = UGC~7436} (Fig~\ref{figvcc0543}). This flattened
rotationally supported galaxy was a subject to the several kinematical studies
referenced above. It is one of the three galaxies with the data available
both in the Palomar dE project and in the HFA. The kinematical and stellar
population profiles are consistent between the datasets. The age
profile in this object is nearly flat with a mean value of
$\sim$5.5~Gyr, while the metallicity exhibits a slight gradient. No embedded
structures were detected in this object by \cite{LGB06}.

\noindent\emph{VCC~545 = IC~783A} No substructures were detected by \cite{LGB06} in
this faint nucleated dE galaxy. The signal-to-noise ratio of the HFA data
was very low so we were able to determine only integrated properties of the
stellar populations and mean velocity dispersion, which agree well with the
values derived from fitting the SDSS DR6 spectrum. The galaxy is located
close in projection to VCC~490 and M~100 and its radial velocity is quite
close to that of VCC~490.

\begin{figure}
\includegraphics[width=\hsize]{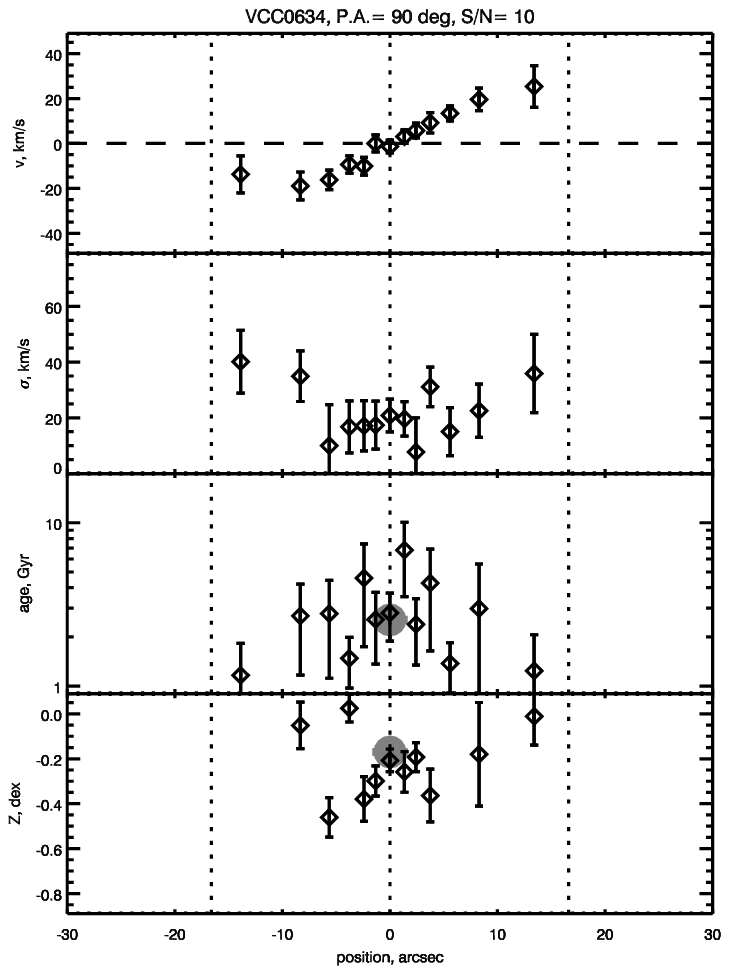}
\caption{Kinematics and stellar populations of VCC~634 (OHP CARELEC).
See Fig~\ref{figvcc0178} for details. \label{figvcc0634}}
\end{figure}

\noindent\emph{VCC~634 = NGC~4328} (Fig~\ref{figvcc0634}). This relatively
bright rotationally supported nucleated dwarf galaxy does not exhibit
embedded structures \citep{LGB06}. We see the intermediate-age (3~Gyr) quite
metal-rich ($-0.2$~dex) stellar population without significant changes of the
properties within 1~$r_{e}$. Low signal-to-noise ratio of the data did not
allow us to measure the metallicity behaviour at larger radii.

\begin{figure}
\includegraphics[width=\hsize]{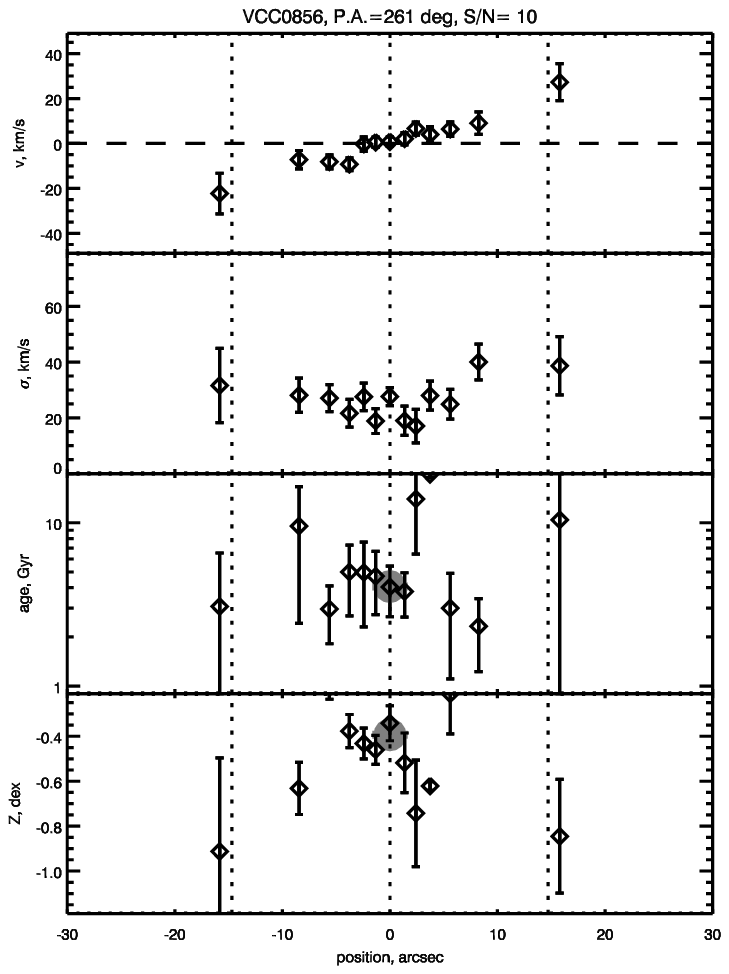}
\caption{Kinematics and stellar populations of VCC~856 (OHP CARELEC).
See Fig~\ref{figvcc0178} for details. \label{figvcc0856}}
\end{figure}

\noindent\emph{VCC~856 = IC~3328} (Fig~\ref{figvcc0856}). This was the first
dwarf dE/dS0 galaxy, where low-contrast spiral arms were detected by
\cite{JKB00}. The data allowed us to derive the stellar population
properties out to 1~$r_{e}$. The age profile is flat with the mean value of
about 4~Gyr, whereas the metallicity distribution exhibits a gradient. With
the presently available data we were not able to detect any statistically
significant difference between the ages of the stellar populations in the
nuclear region and in the spiral disc.

\begin{figure}
\includegraphics[width=\hsize]{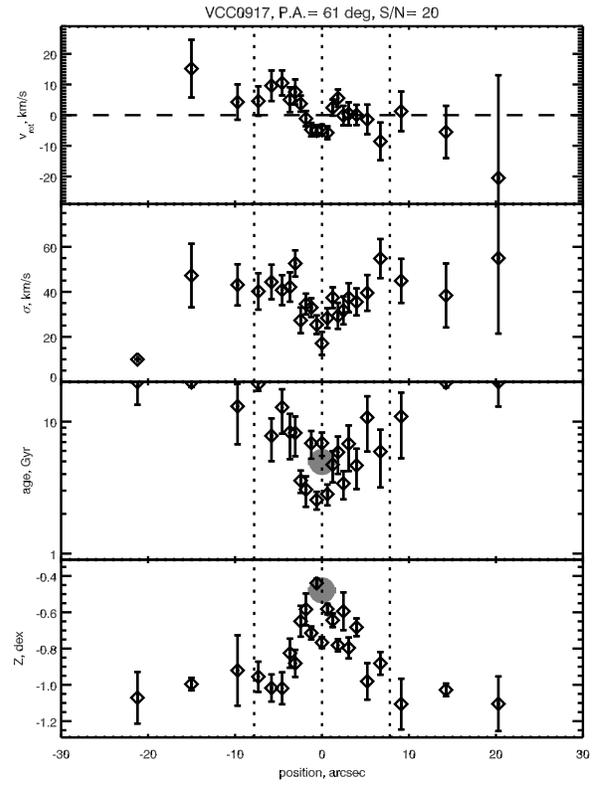}
\caption{Kinematics and stellar populations of VCC~917 (Palomar DS).
See Fig~\ref{figvcc0178} for details. \label{figvcc0917}}
\end{figure}

\noindent\emph{VCC~917 = IC~3344} (Fig~\ref{figvcc0917}). No substructures were
revealed by \cite{LGB06} in VCC~917. This flattened object with a very
little rotation was considered as one of the examples of the galaxies
supported by the anisotropic velocity dispersions. We see a kinematically
decoupled central component with the radius of about 3~arcsec
($\sim$0.25~kpc) rotating in the other sense compared to the main galactic
body. It is associated with the drop in the velocity dispersion profile from
$\sim$40 to $\sim$20~km~s$^{-1}$ and younger and much more metal-rich
stellar population compared to the periphery. This is the first example of
the KDC in a dwarf galaxy supported by anisotropic velocity dispersions.

\begin{figure}
\includegraphics[width=\hsize]{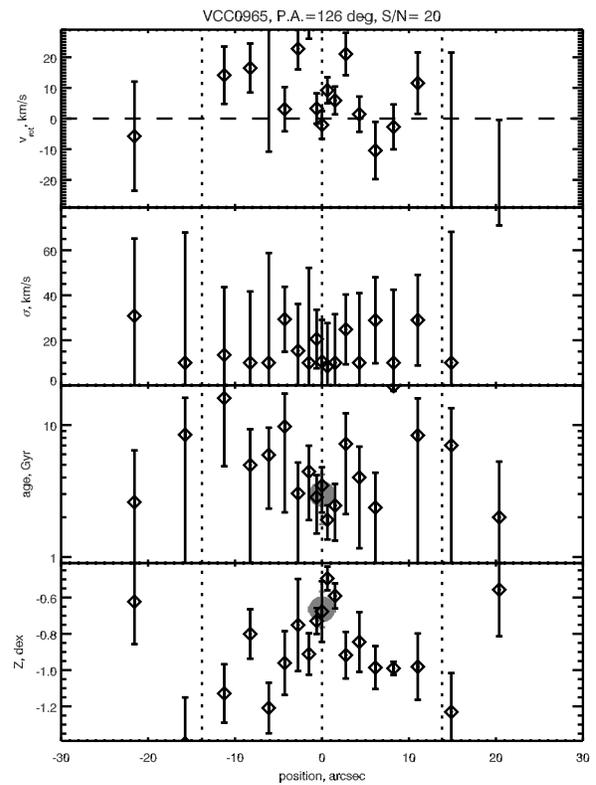}
\caption{Kinematics and stellar populations of VCC~965 (Palomar DS).
See Fig~\ref{figvcc0178} for details. \label{figvcc0965}}
\end{figure}

\noindent\emph{VCC~965 = IC~3363} (Fig~\ref{figvcc0965}). No substructures are
reported by \cite{LGB06} in this faint flattened nucleated galaxy. This is
one of a few cases, where the internal velocity dispersion is too low to be
measured. The stellar population in the nuclear region turns to be very
different from the periphery of the galaxy both in terms of young age and
high metallicity. The velocity dispersion profile of this object derived
from the red arm DS spectra and presented in \cite{vZSH04} demonstrates a
central velocity dispersion drop.

\begin{figure}
\includegraphics[width=\hsize]{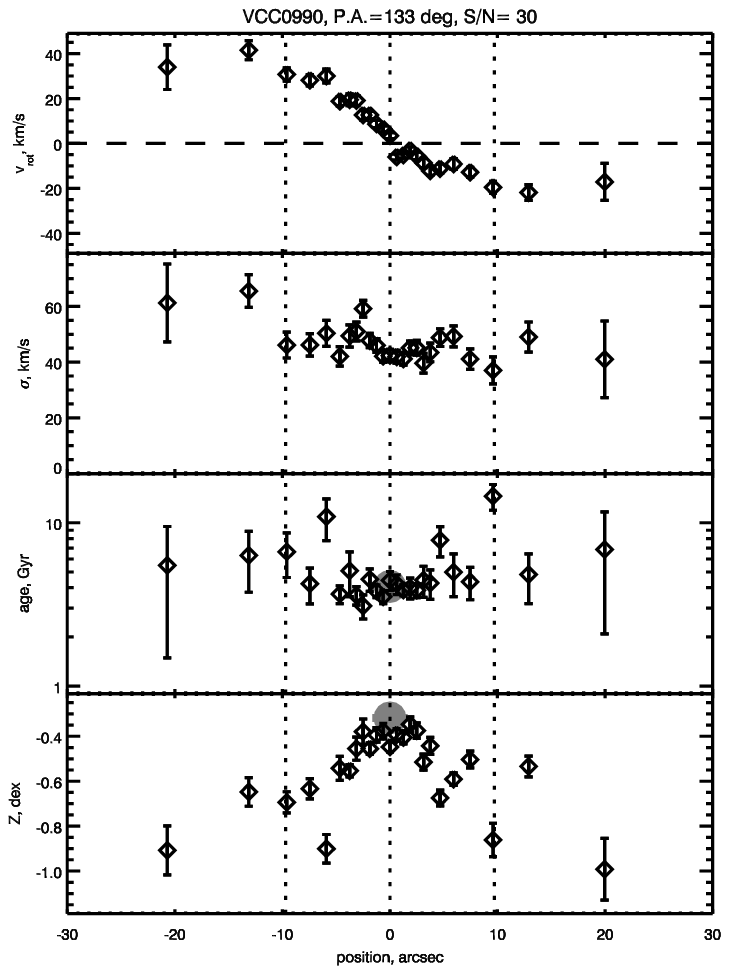}
\caption{Kinematics and stellar populations of VCC~990 (Palomar DS).
See Fig~\ref{figvcc0178} for details. \label{figvcc0990}}
\end{figure}

\noindent\emph{VCC~990 = IC~3369} (Fig~\ref{figvcc0990}). \cite{LGB06} report
the detection of an inclined disc in this flattened nucleated galaxy. Inside
1~$r_{e}$ we see a quite strong metallicity gradient from about $-0.3$~dex to
$-0.8$~dex while the age remains nearly constant at a level of $\sim$5~Gyr.
The velocity dispersion distribution exhibits a local minimum in the centre
associated with a peculiarity on the radial velocity profile. We detected
faint emission lines (H$\beta$, [OIII]) in the fitting residuals suggesting
the presence of ionised gas in this galaxy.

\begin{figure}
\includegraphics[width=\hsize]{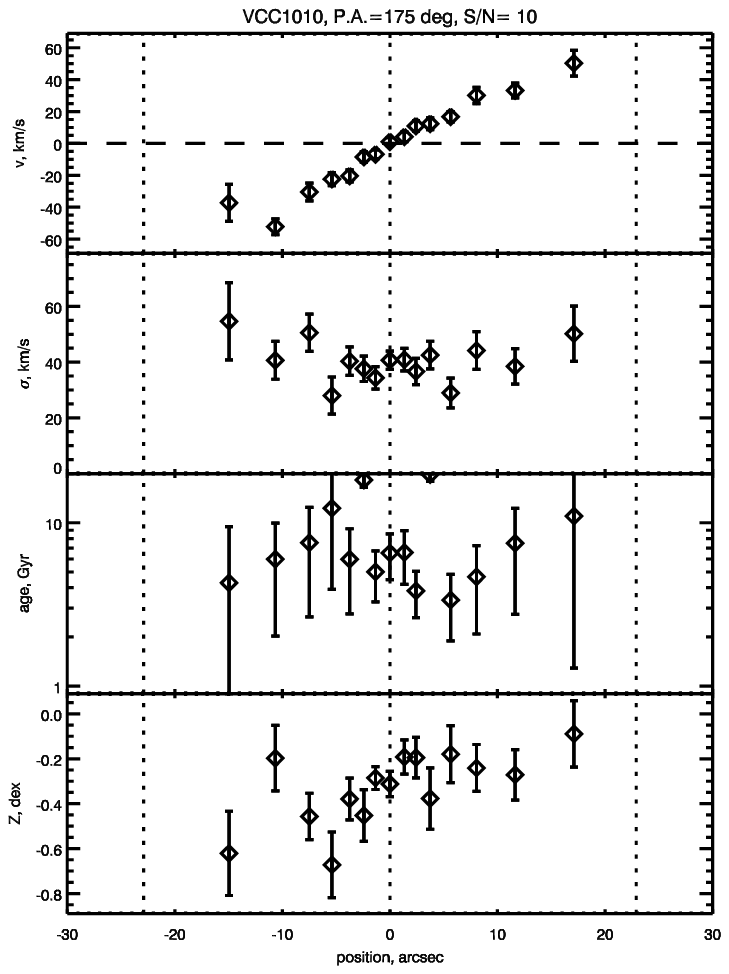}
\caption{Kinematics and stellar populations of VCC~1010 (OHP CARELEC).
See Fig~\ref{figvcc0178} for details. \label{figvcc1010}}
\end{figure}

\noindent\emph{VCC~1010 = NGC~4431} (Fig~\ref{figvcc1010}). Bar and disc were
detected in this flattened rotationally supported galaxy by \cite{LGB06}.
The data available for this object in the HFA allowed us to trace the
behaviour of the stellar population parameters out to $r_{e} / 2$. Both, age
and metallicity distribution look flat showing no significant changes along
the radius.

\begin{figure}
\includegraphics[width=\hsize]{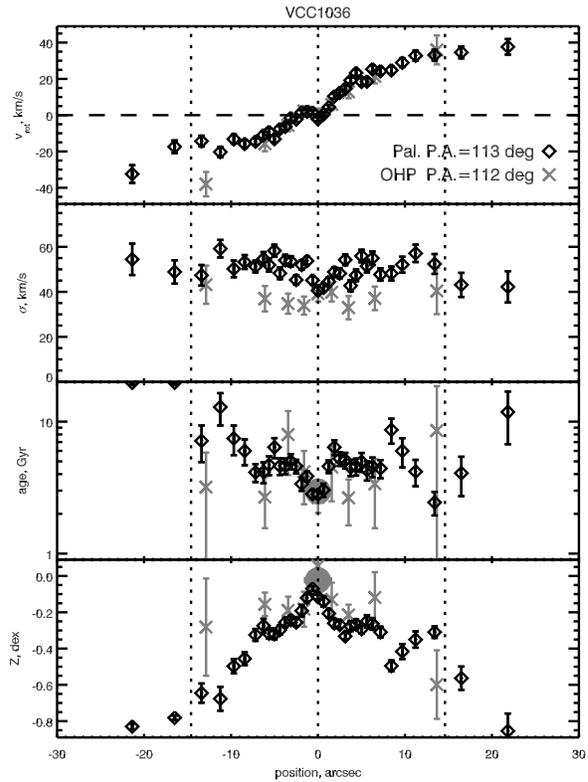}
\caption{Kinematics and stellar populations of VCC~1036 (Palomar DS and OHP
CARELEC). The target signal-to-noise ratios were 30 and 10 for the Palomar
DS and OHP datasets respectively. See Fig~\ref{figvcc0543} for details. 
\label{figvcc1036}}
\end{figure}

\noindent\emph{VCC~1036 = NGC~4436} (Fig~\ref{figvcc1036}). The inclined
disc is reported by \cite{LGB06}. Nearly solid-body rotation out to 1~$r_e$
changes into the flattening of the rotation curve outwards. The velocity
dispersion stays constant beyond 0.3~$r_e$ at a level of
$\sim$50~km~s$^{-1}$. This galaxy exhibits a prominent KDC. It is one of the
three objects included both in the Palomar dE sample and in the sample of
\cite{SP02}. Both datasets agree well on the presence of the kinematically
decoupled central component. However, due to worse seeing conditions it is
not seen as counter rotation in the HFA data, but as a flat region of the
radial velocity profile. A slight depression is seen in the velocity
dispersion profile in the region corresponding to the KDC. It corresponds to
a pronounced bump in the metallicity profile with the values reaching almost
[Fe/H]=0.0 in the SDSS data. SSP-equivalent age ($\sim$3Gyr) in this part
is $\sim$2~Gyr younger than in the surrounding regions. The metallicity keeps
constant at a level about $-0.3$~dex out to 8~arcsec, then it starts to
decrease steeply. This is the only object, where the central value of the
$\alpha$/Fe abundance ratio ($-0.05$~dex) differs strongly from the values
in the periphery ($+0.17$~dex).

\begin{figure}
\includegraphics[width=\hsize]{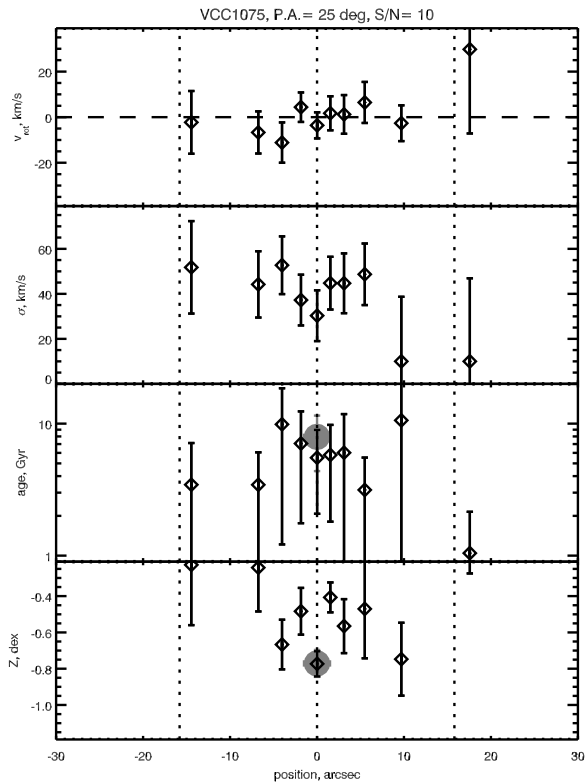}
\caption{Kinematics and stellar populations of VCC~1075 (Palomar DS).
See Fig~\ref{figvcc0178} for details. \label{figvcc1075}}
\end{figure}

\noindent\emph{VCC~1075 = IC~3383} (Fig~\ref{figvcc1075}). This faint flattened
nucleated dwarf galaxy harbours no embedded substructures according to
\cite{LGB06}. The signal-to-noise ratio of the data is quite low, making
difficult to assess the variations of the stellar population parameters
along the radius. However, at the present level of precision we do not
detect any statistically significant gradients of age neither metallicity.

\begin{figure}
\includegraphics[width=\hsize]{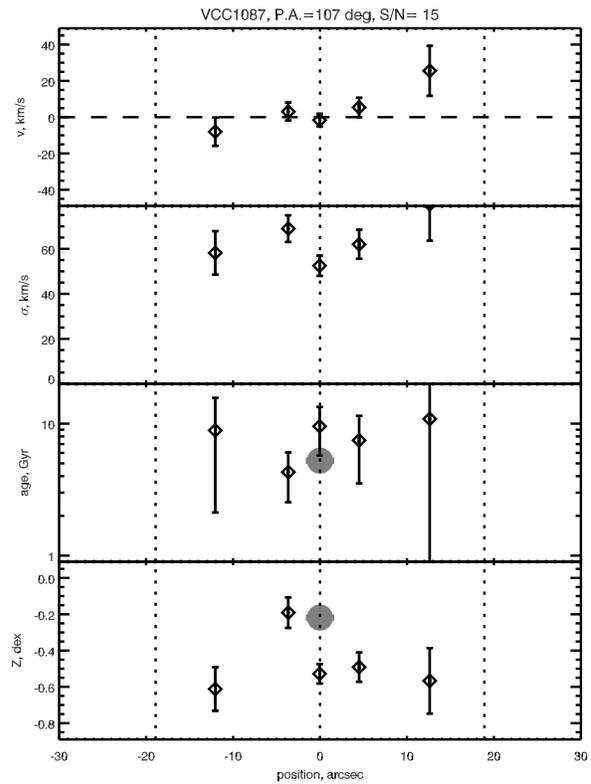}
\caption{Kinematics and stellar populations of VCC~1087 (OHP CARELEC).
See Fig~\ref{figvcc0178} for details. \label{figvcc1087}}
\end{figure}

\noindent\emph{VCC~1087 = IC~3381} (Fig~\ref{figvcc1087}). This flattened
nucleated dwarf with no substructures reported by \cite{LGB06} has very
little if any rotation. The low signal-to-noise ratio of the data makes
difficult to study the variations of the stellar population along the
radius. We notice that the stellar population parameters derived from the
SDSS DR6 spectrum fitting are very different from those estimated from the
HFA spectra. This, however, may be explained as the seeing effect, which
smoothed the nucleus.

\begin{figure}
\includegraphics[width=\hsize]{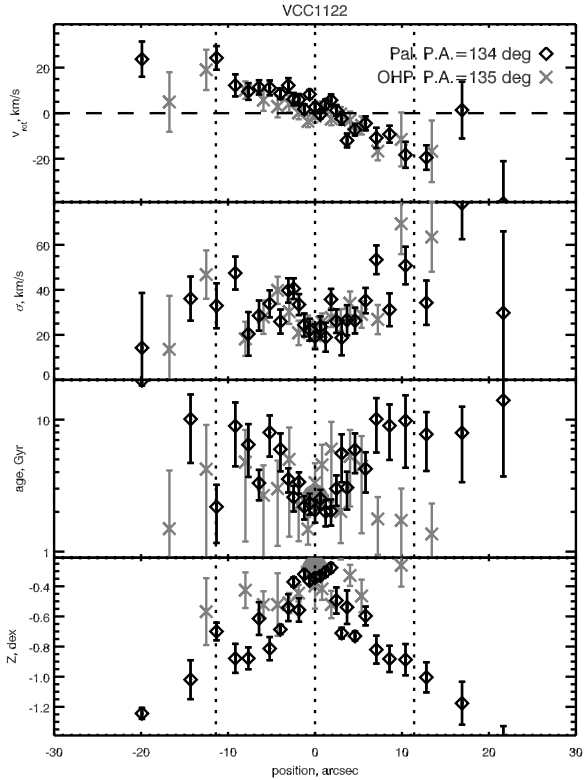}
\caption{Kinematics and stellar populations of VCC~1122 (Palomar DS and OHP
CARELEC). The target signal-to-noise ratios were 30 and 10 for the Palomar  
DS and OHP datasets respectively. See Fig~\ref{figvcc0543} for details.
\label{figvcc1122}}
\end{figure}

\noindent\emph{VCC~1122 = IC~3393} (Fig~\ref{figvcc1122}). This strongly
flattened galaxy with the insufficient degree of rotation to be considered
as rotationally supported does not exhibit embedded substructures according
to \cite{LGB06}. We have the data from both Palomar dE project and the HFA.
The kinematical profiles agree quite well. There is a peculiarity seen in
the radial velocity profile 2~arcsec south-east of the nucleus, however
presently we do not have enough evidences to interpret it as a kinematically
decoupled component. The stellar population properties were determined out
to 2~$r_{e}$ from the Palomar dE data. The nuclear region of VCC~1122 is
clearly distinct in both age and metallicity profiles. We also see a
remarkable metallicity gradient, which is, however still insufficient to
explain the high metallicity in the nuclear region. The age stays constant
along the radius beyond $r_{e} / 2$. There are faint H$\beta$ and [OIII]
emission lines revealed in the fitting residuals localised in the central
part of the galaxy.

\begin{figure}
\includegraphics[width=\hsize]{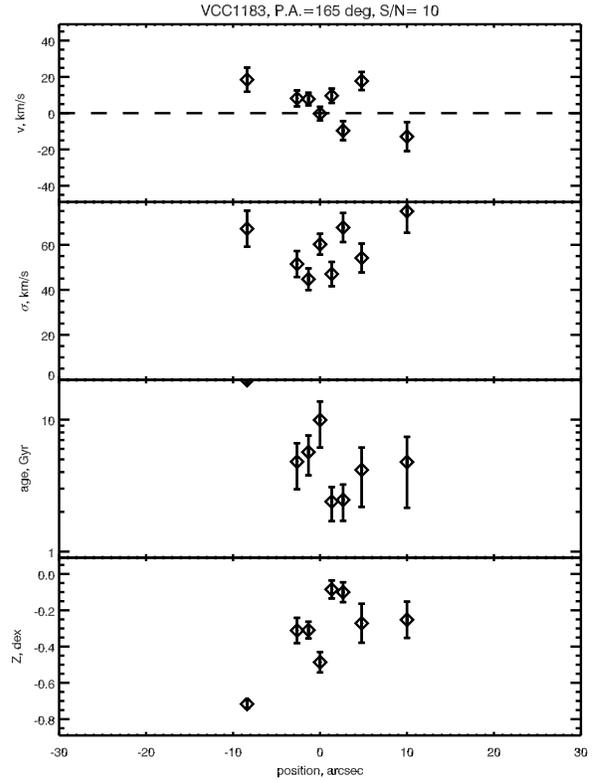}
\caption{Kinematics and stellar populations of VCC~1183 (OHP CARELEC).
See Fig~\ref{figvcc0178} for details. \label{figvcc1183}}
\end{figure}

\noindent\emph{VCC~1183 = IC~3413} (Fig~\ref{figvcc1183}). This flattened
nucleated dE galaxy harbours a bar reported by \cite{LGB06}. We observe some
rotation. Low signal-to-noise ratio of the data did not allow the detailed
studies of the stellar populations, however we notice the intermediate age
($\sim$4~Gyr) and relatively high metallicity ($-0.3$~dex) without evident
changes along the radius.

\begin{figure}
\includegraphics[width=\hsize]{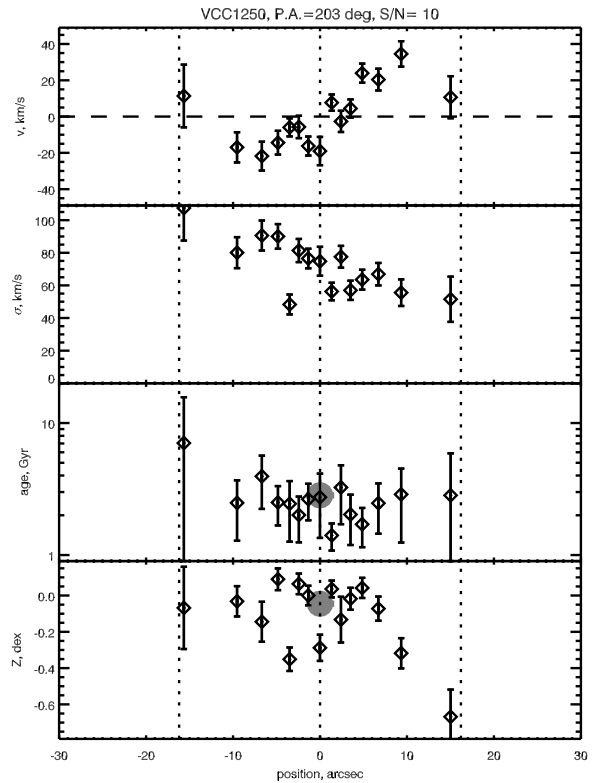}
\caption{Kinematics and stellar populations of VCC~1250 (OHP CARELEC).
See Fig~\ref{figvcc0178} for details. \label{figvcc1250}}
\end{figure}

\noindent\emph{VCC~1250 = NGC~4476} (Fig~\ref{figvcc1250}). This galaxy hosts a
large embedded disc with the spiral arms, dust lanes, and HII regions
clearly visible on the HST ACS images presented in \cite{Ferrarese+06}. Our
data covers the region completely dominated by the disc. We see prominent
emission lines suggesting on-going star formation. The luminosity weighted
age is about 2~Gyr and the metallicity is as high as solar without
statistically significant changes along the radius. The $B$-band absolute
magnitude of VCC~1250, $-18.27$~mag places it at the limit of the informal
dwarf/giant separation. However, it still can be classified as a
low-luminosity early-type galaxy, that's why we included it into our sample.

\begin{figure}
\includegraphics[width=\hsize]{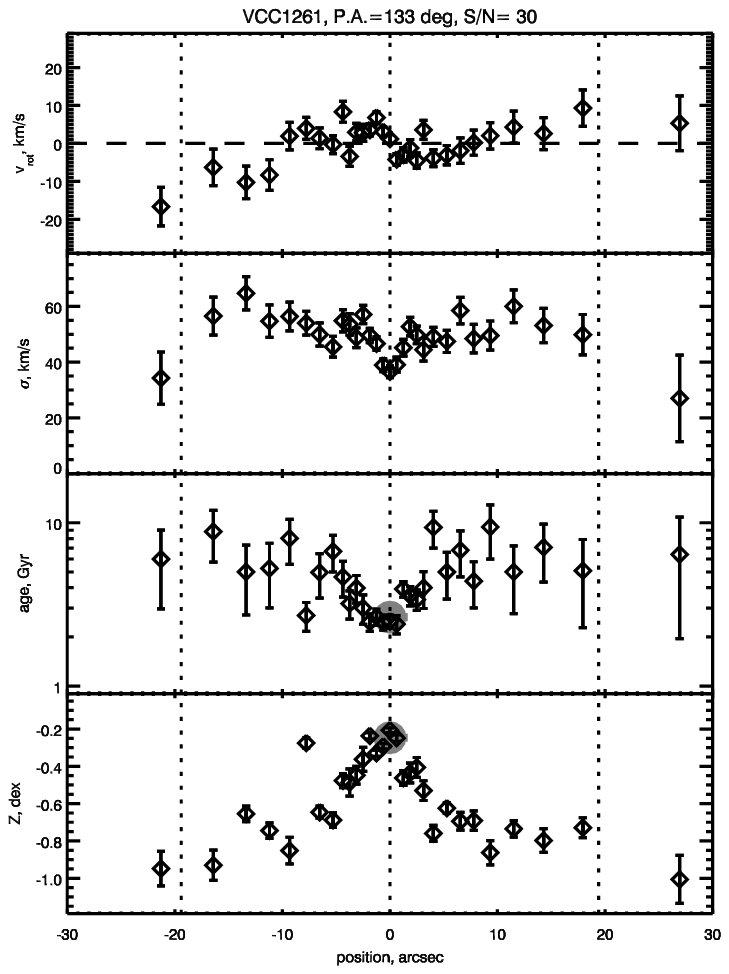}
\caption{Kinematics and stellar populations of VCC~1261 (Palomar DS).
See Fig~\ref{figvcc0178} for details. \label{figvcc1261}}
\end{figure}

\noindent\emph{VCC~1261 = NGC~4482} (Fig~\ref{figvcc1261}). No substructures
were reported by \cite{LGB06} in this galaxy. This object looks like an
enlarged version of VCC~917, giving another example of a small rapidly
rotating KDC in a flattened spheroid with a very low, if any, large-scale
rotation. As in VCC~917, the KDC is associated with a velocity dispersion
drop, younger and much more metal rich stellar population than in the
peripheral parts of the galaxy. Faint emission lines are seen in the fitting
residuals, suggesting the presence of warm ISM in this object.

\begin{figure}
\includegraphics[width=\hsize]{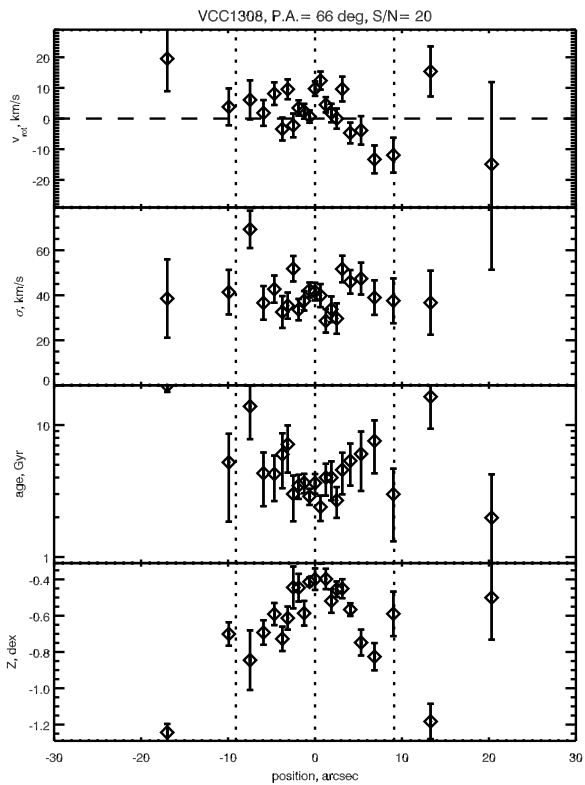}
\caption{Kinematics and stellar populations of VCC~1308 (Palomar DS).
See Fig~\ref{figvcc0178} for details. \label{figvcc1308}}
\end{figure}

\noindent\emph{VCC~1308 = IC~3437} (Fig~\ref{figvcc1308}). This galaxy was
classified as ``non-rotating'' by \cite{GGvdM03} but as ``rotating'' by
\cite{vZSH04}. No substructures were reported in it by \cite{LGB06}. We
indeed see some rotation, although the radial velocity profile is quite
peculiar in the centre, as well as a remarkable metallicity gradient. The
stellar population in the central region of the galaxy looks younger then in
the outer regions along the radius.

\noindent\emph{VCC~1407 = IC~3461} Very low signal-to-noise ratio of the
data does not allow to study the variations of the stellar population properties
along the radius. However, we stress a very good agreement of the age and
metallicity estimates for this object obtained from the fitting of the HFA
and SDSS DR6 spectra. No embedded structures were reported in this galaxy by
\cite{LGB06}.

\begin{figure}
\begin{tabular}{cc}
\includegraphics[width=0.5\hsize]{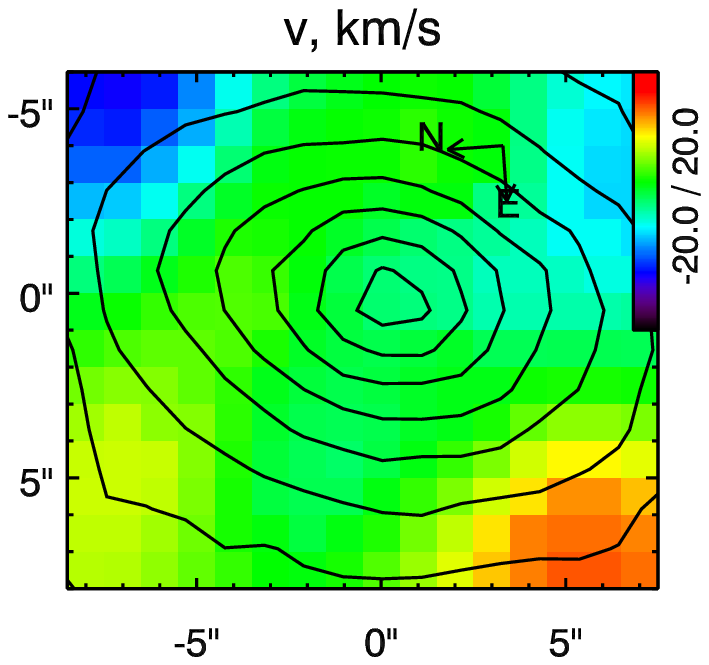} & 
\includegraphics[width=0.5\hsize]{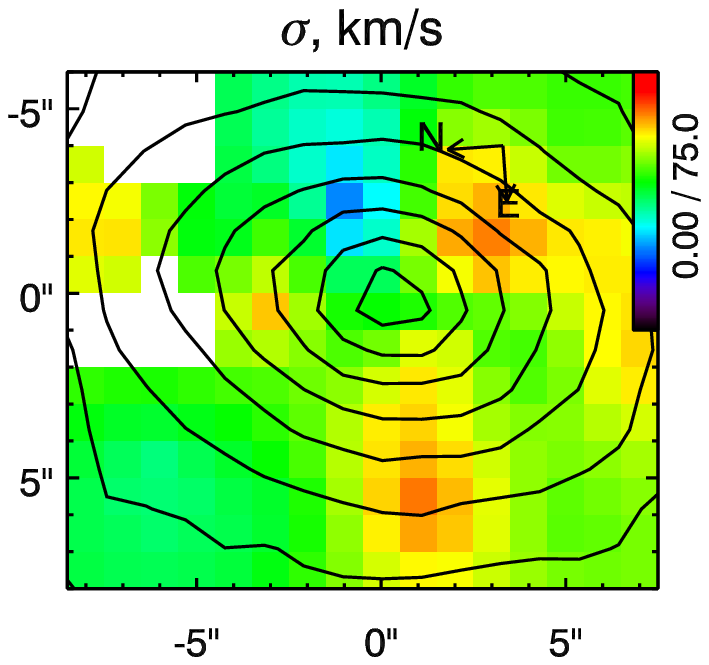} \\
\includegraphics[width=0.5\hsize]{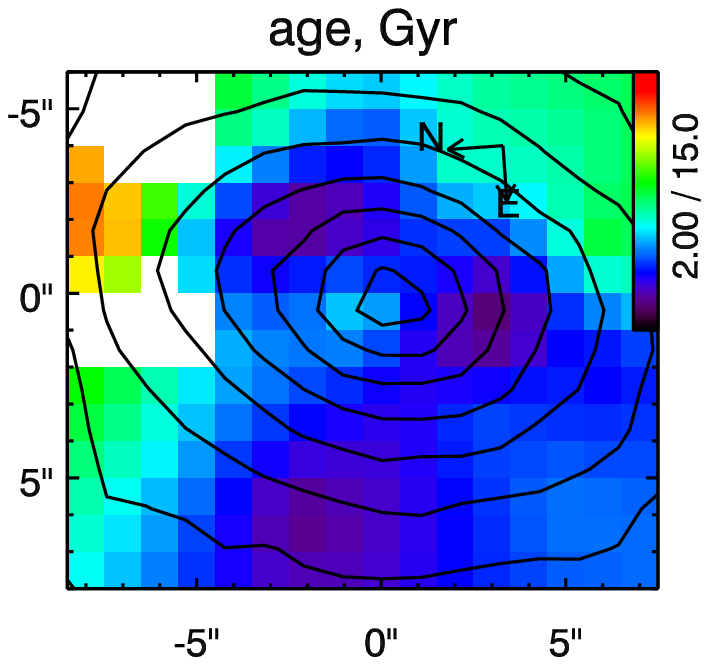} & 
\includegraphics[width=0.5\hsize]{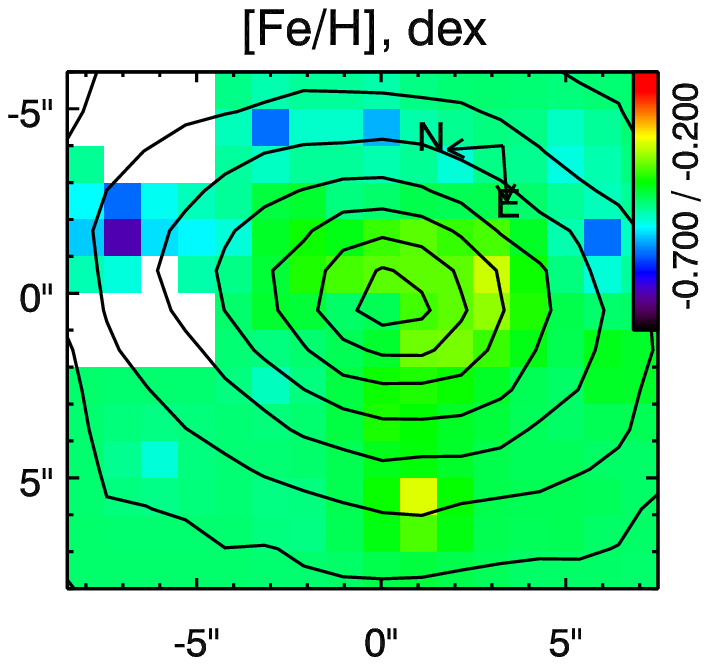} \\
\end{tabular}
\caption{Kinematics and stellar populations of VCC~1422 (MPFS).
Panels are the same as in Fig~\ref{figvcc0490}. 
Adaptive binning with the target signal-to-noise ratio of 15
was applied.\label{figvcc1422}}
\end{figure}

\noindent\emph{VCC~1422 = IC~3468} (Fig~\ref{figvcc1422}). An embedded
structure is detected in this galaxy interpreted as a bar by \cite{BBJ02} or
as an edge-on disc by \cite{LGB06}. The IFU spectroscopy reveals complex
kinematics with evidences for rotation only beyond 5~arcsec from the centre.
There is a hint that stellar population in the region of the disc/bar is
somewhat younger than in the periphery of the galaxy ($\sigma = 54 \pm
4$~km~s~$^{-1}$, $t=7.2$~Gyr, [Fe/H]=-0.54~dex), although the difference is
on the limit of detection. VCC~1422 is the only galaxy in our sample, where
the very central region, dominated by the light of its nucleus exhibits
older and more metal-poor stellar population than one of the surrounding
disc/bar clearly visible on the maps. Its age and metallicity are found to
be very close to those in the periphery of the galaxy.

\begin{figure}
\includegraphics[width=\hsize]{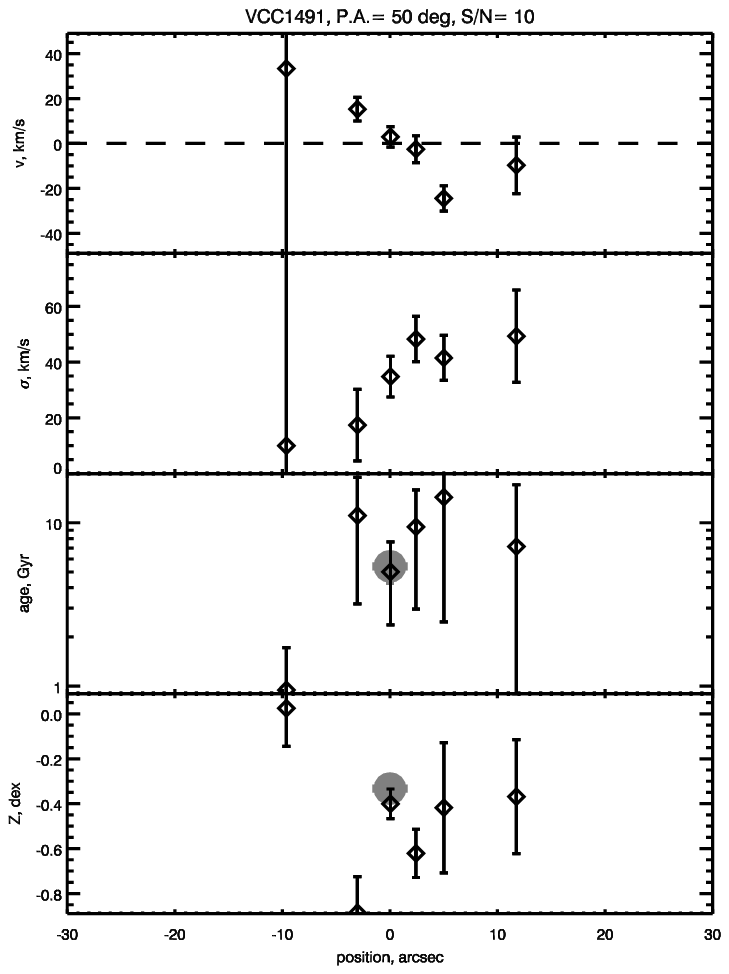}
\caption{Kinematics and stellar populations of VCC~1491 (OHP CARELEC).
See Fig~\ref{figvcc0178} for details. \label{figvcc1491}}
\end{figure}

\noindent\emph{VCC~1491 = IC~3486} (Fig~\ref{figvcc1491}). No substructures
were revealed by \cite{LGB06} in this non-nucleated galaxy demonstrating
quite high degree of rotation for its moderate flattening. The stellar
population is old ($\sim$8~Gyr) and metal-poor ($-0.6$~dex) without noticeable
gradients.

\begin{figure}
\includegraphics[width=\hsize]{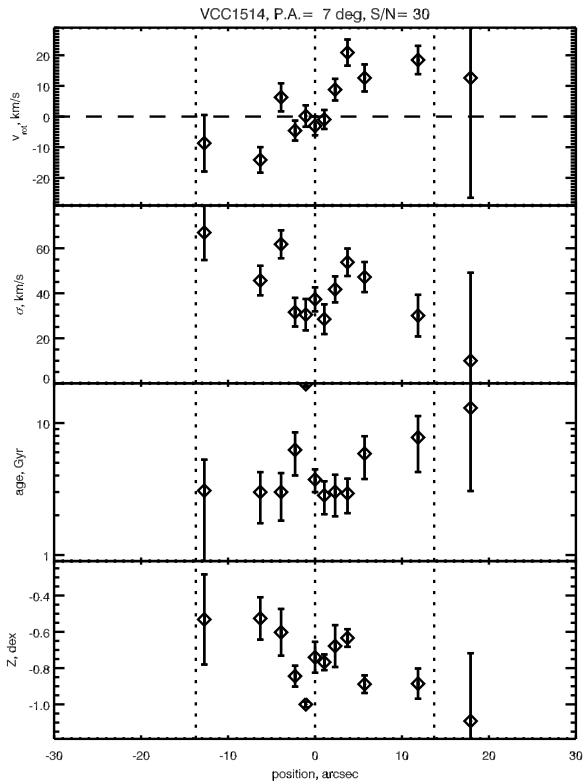}
\caption{Kinematics and stellar populations of VCC~1514 (Palomar DS).
See Fig~\ref{figvcc0178} for details. \label{figvcc1514}}
\end{figure}

\noindent\emph{VCC~1514 = PGC~41726} (Fig~\ref{figvcc1514}). Possible inclined
disc was reported by \cite{LGB06} in this strongly flattened galaxy. We
observe some rotation. The age and metallicity estimates do not exhibit
statistically significant variations along the radius, although the
signal-to-noise ratio of the data was rather low.

\begin{figure}
\begin{tabular}{cc}
\includegraphics[width=0.5\hsize]{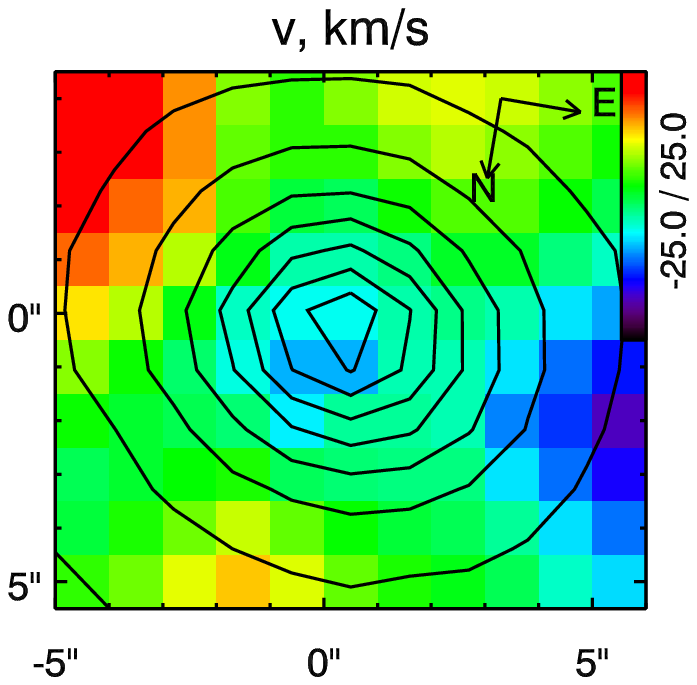} & 
\includegraphics[width=0.5\hsize]{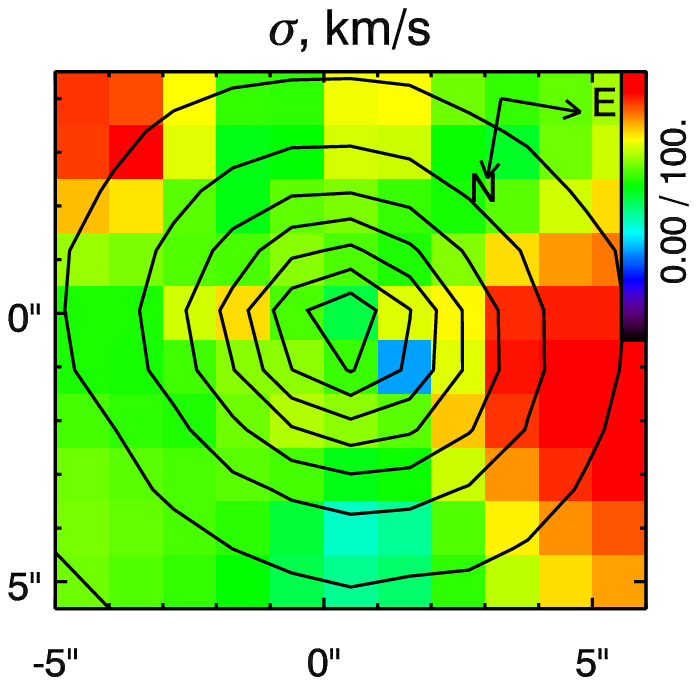} \\
\includegraphics[width=0.5\hsize]{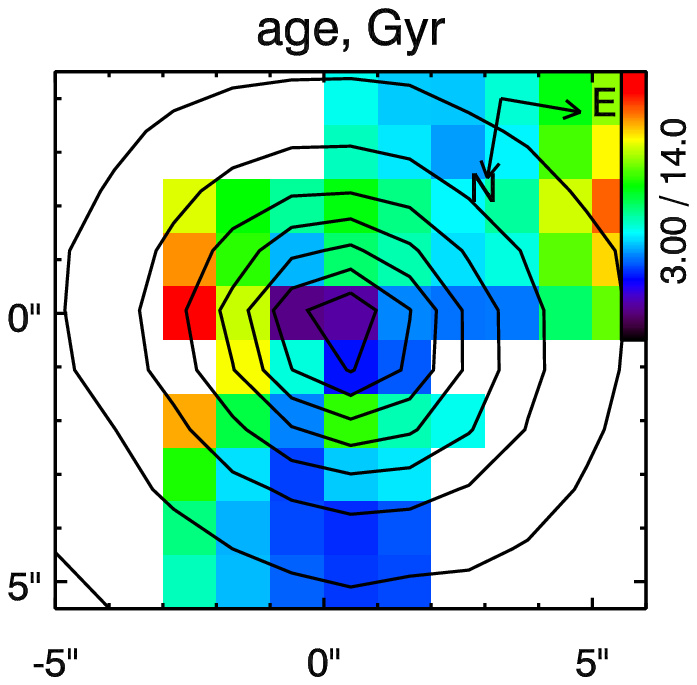} & 
\includegraphics[width=0.5\hsize]{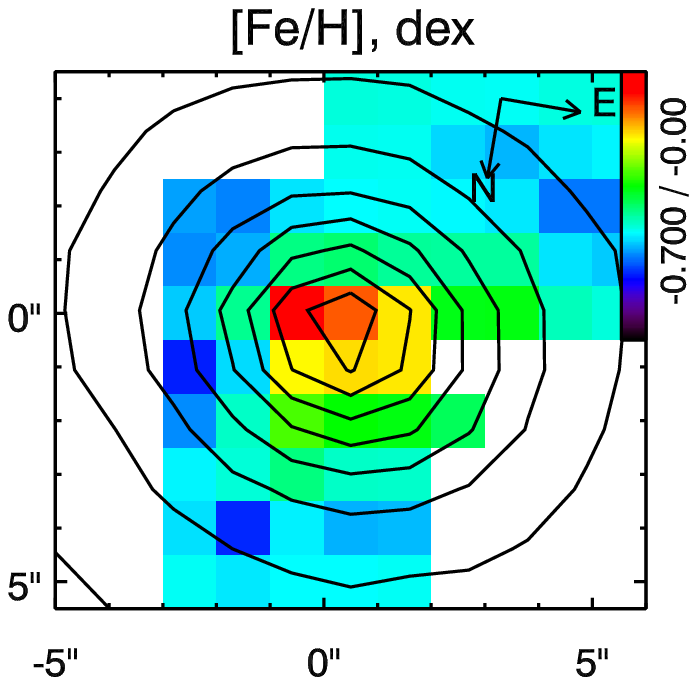} \\
\end{tabular}
\caption{Kinematics and stellar populations of VCC~1545 (MPFS).
Panels are the same as in Fig~\ref{figvcc0490}. 
Adaptive binning with the target signal-to-noise ratio of 15
was applied.\label{figvcc1545}}
\end{figure}

\noindent\emph{VCC~1545 = IC~3509} (Fig~\ref{figvcc1545}). This galaxy was
among the first, where the young stellar population in the central region
was reported by \cite{CSAP07}. It demonstrates very complex kinematics: we
see a minor-axis rotation (i.e. kinematical decoupling) in the circumnuclear
region of the galaxy with the depression in the velocity dispersion
distribution there, while there is a large-scale major-axis rotation as
well. The stellar population in the nuclear region is younger and
considerably more metal-rich than in the main galactic body.

\noindent\emph{VCC~1743 = IC~3602} This is a faint flattened dwarf galaxy
with the faint extended nucleus seen on the HST ACS images but not on the
ground-based data. The galaxy is listed in Appendix~E of of \cite{LGB06} as
an object ``in which substructure other than a disc was found''. The data
available to us have very low signal-to-noise ratio due to low surface
brightness of the galaxy, therefore we were not able to derive detailed
profiles of the stellar population properties. However, outside the nuclear
region the age looks to be old ($\sim$10~Gyr) and the metallicity quite low
($-0.6$~dex).

\noindent\emph{VCC~1857 = IC~3647} The faintest object in the Palomar dE
sample does not exhibit and substructures reported by \cite{LGB06}. The
signal-to-noise is not sufficient even for the precise integrated measurements
of the stellar populations, however at 1-$\sigma$ confidence level the
galaxy has and intermediate age not older than 7~Gyr.

\noindent\emph{VCC~1871 = IC~3653} An inclined stellar disc was discovered
in the velocity field by \cite{CPSA07} with a counterpart in the colour map.
The galaxy has somewhat high surface brightness, the age distribution is
nearly uniform, whereas there is a pronounced peak in the metallicity map
with the the values slightly exceeding solar.

\begin{figure}
\includegraphics[width=\hsize]{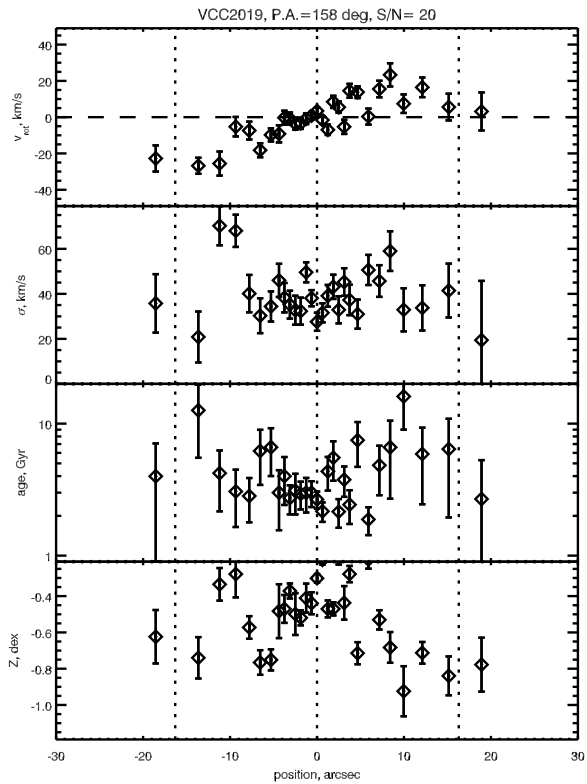}
\caption{Kinematics and stellar populations of VCC~2019 (Palomar DS).
See Fig~\ref{figvcc0178} for details. \label{figvcc2019}}
\end{figure}

\noindent\emph{VCC~2019 = IC~3735} (Fig~\ref{figvcc2019}). A ``possible
inclined disc, maybe warped or distorted'' is reported in this galaxy by
\cite{LGB06}. The isophotes demonstrate the change of ellipticity and
positional angle \citep{Ferrarese+06}. We see a central peak in the
metallicity distribution and the corresponding drop in the velocity
dispersion profile, while the quality of the age measurements is not
sufficient to clearly say whether the nucleus harbours younger stellar
population than the outer parts of the galaxy. A noticeable metallicity
gradient is observed. We detected faint emission lines (H$\beta$ and [OIII])
in the fitting residuals.

\begin{figure}
\includegraphics[width=\hsize]{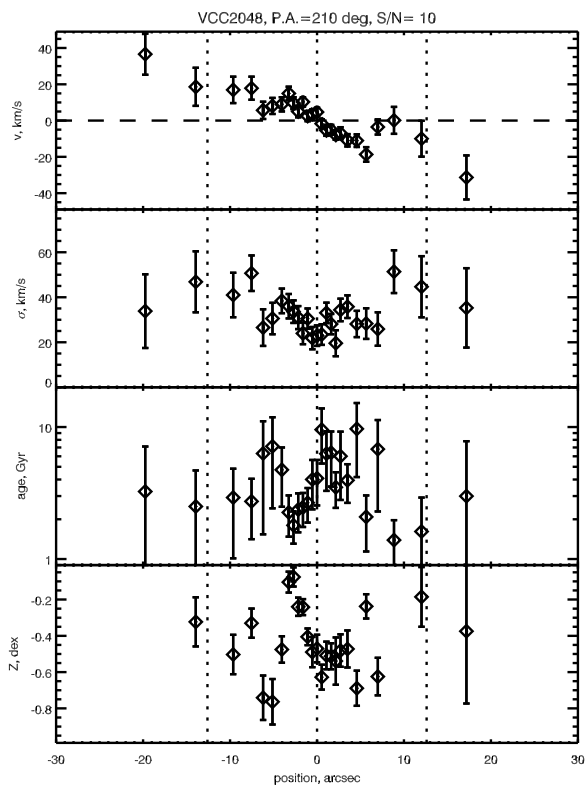}
\caption{Kinematics and stellar populations of VCC~2048 (OHP CARELEC).
See Fig~\ref{figvcc0178} for details. \label{figvcc2048}}
\end{figure}

\noindent\emph{VCC~2048 = IC~3773} (Fig~\ref{figvcc2048}). There is an evidence
for a bar or edge-on embedded disc reported by \cite{LGB06}. This nucleated
galaxy is strongly flattened. Its inner region, corresponding to the disc
demonstrates fast solid-body rotation and looks kinematically decoupled from
the outer parts of the galaxy rotating in the same sense, similar to the
structures observed by \cite{DRDZH04} in the two dE galaxies in groups.
There is a pronounced $\sigma$-drop in the central region of the galaxy.
Stellar population properties do not show statistically significant changes
along the radius apart from the little region 4~arcsec North of the nucleus,
where the metallicity rises by $\sim$0.3~dex compared to the nearby regions.
Presently we cannot explain this feature, it may be connected to the
globular cluster which might have fallen into the slit.

\begin{figure}
\includegraphics[width=\hsize]{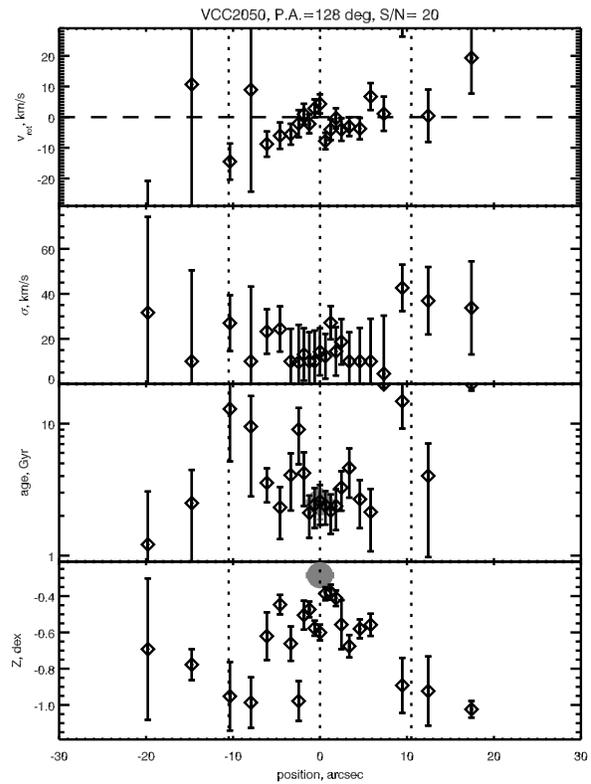}
\caption{Kinematics and stellar populations of VCC~2050 (Palomar DS).
See Fig~\ref{figvcc0178} for details. \label{figvcc2050}}
\end{figure}

\noindent\emph{VCC~2050 = IC~3779} (Fig~\ref{figvcc2050}). Probable inclined
disc is reported in this galaxy by \cite{LGB06}. The observed velocity
dispersion in this object is too low to be precisely measured. There is a
significant metallicity gradient, whereas the signal is not sufficient to
detect any changes in the age distribution. The luminosity-weighted age in
the central part is quite young ($\sim$2.5~Gyr) and we see faint emission
lines in the fitting residuals.

\section{Summary and Conclusions}

We have presented the first large dataset of spectroscopically derived
radial profiles of stellar population parameters for dwarf early-type
galaxies made by re-analysing some of the published spectroscopic data.
Thanks to the usage of the {\sc NBursts} full spectral fitting technique we
have improved the stellar kinematics compared to \cite{vZSH04} reaching
comparable quality of measurements to those presented in
\cite{GGvdM02,GGvdM03} but generally going out to 1~$r_e$ or further from
the centre. We have performed the stellar population analysis of the spectra
where only kinematical data were available \citep{SP02}. Our measurements of
the nuclear stellar populations are in a good agreement with the results
obtained from the fitting of SDSS DR6 spectra.

Significant metallicity gradients are often observed in dE/dS0 galaxies,
although in some of the objects the metallicity distribution is completely
constant along the radius. There is a tendency to the dichotomy in the
distribution of the metallicity gradients -- metallicity profiles are either
nearly flat or steeply decreasing along the radius, reaching
$-0.9$~dex~$r_e^{-1}$ in VCC~437. In none of the cases we see a statistically
significant positive gradient. Age distribution along the radius usually
exhibits flat behaviour. If any gradients in age exist, they cannot be
clearly detected using the existing datasets.

Levels of $\alpha$-enhancement as derived from the Lick indices exhibit no
statistically significant changes along the radius and between the nuclei
and main discs/spheroids in all objects, but VCC~1036, where the value in
the nuclear region is nearly solar ($[$Mg/Fe$]$=$-0.05$~dex), whereas it
stays super-solar (+0.17~dex) along the radius out to 1.2~$r_e$.

Stellar populations in the nuclear regions are usually more metal-rich than
in the main galactic bodies, and the central values of the
luminosity-weighted metallicity are higher than what one would expect from
the simple interpolation of the metallicity gradients toward the centres. A
remarkable fraction of galaxies (11 or 27) demonstrate luminosity-weighted
ages in the nuclear regions to be younger than those of the main galactic
bodies, with the difference exceeding 7~Gyr in VCC~917 and VCC~490.

These chemically and evolutionary decoupled nuclear regions are always
associated with the drops in the velocity dispersion profiles. Only in those
cases, where the velocity dispersions are too low to be measured with the
present data, the $\sigma$-drops are not observed. In these cases, the
mass-to-light ratios change significantly along the radius, pointing out
to unrighteous use of constant $M/L$ dynamical models.

In certain objects (VCC~990, 1122, 1250, 1261, 2019, and 2050) we detected
faint emission lines (H$\beta$, [OIII]) in the nuclear regions, associated
with the chemically and evolutionary decoupled cores.

We discovered four kinematically decoupled rotating cores in bright members
of our sample (VCC~917, 1036, 1261, and 1545), two of them (VCC~917 and
1261) being hosted by the galaxies with very low if any major axis rotation.
In all cases KDCs are associated with young metal-rich stellar populations
and velocity dispersion drops.

Finally, we see remarkable resemblance between internal structure of dE/dS0s
and of intermediate-mass and giant early-type galaxies. Some low-luminosity
objects after detailed analysis look like ``scaled down'' versions of more
luminous counterparts (e.g. VCC~917/1261 and NGC~5813, VCC~1545 and
NGC~4365), suggesting possible similarities between their evolutionary
paths. Some physical processes, e.g. major mergers and secular evolution,
are usually considered to rule the life of giant early-type galaxies. We
have to revise their applicability to explain the dE/dS0 evolution.

\section*{Acknowledgments}

Author is grateful to Liese Van Zee, who made the Palomar dE data available
making this study possible. Author thanks Gary Mamon, Francoise Combes,
Veronique Cayatte, Paola Di Matteo, and Olga Sil'chenko for useful
discussions and suggestions and to an anonymous referee for valuable
comments which helped to improve this manuscript. This project is supported
by the RFBR grant 07-02-00229-a. This research has made use of SAOImage DS9,
developed by Smithsonian Astrophysical Observatory; Aladin developed by the
Centre de Donn\'ees Astronomiques de Strasbourg; ``exploresdss'' script by
G.~Mamon. We used the spectra provided by the HyperLeda FITS archive and by
the SDSS project.

Funding for the SDSS and SDSS-II has been provided by the Alfred P. Sloan
Foundation, the Participating Institutions, the National Science Foundation,
the U.S. Department of Energy, the National Aeronautics and Space
Administration, the Japanese Monbukagakusho, the Max Planck Society, and the
Higher Education Funding Council for England. The SDSS Web Site is
http://www.sdss.org/.

\bibliographystyle{mn2e}
\bibliography{dE_stpop}

\begin{thebibliography}{}

\bibitem[\protect\citeauthoryear{{Abadi}, {Moore} \& {Bower}}{{Abadi}
  et~al.}{1999}]{AMB99}
{Abadi} M.~G.,  {Moore} B.,    {Bower} R.~G.,  1999, \mnras, 308, 947

\bibitem[\protect\citeauthoryear{{Adelman-McCarthy} et~al.}{{Adelman-McCarthy} et~al.}{2008}]{SDSS_DR6}
{Adelman-McCarthy} J.~K. et al., 2008, \apjs, 175, 297

\bibitem[\protect\citeauthoryear{{Afanasiev}, {Dodonov} \&
  {Moiseev}}{{Afanasiev} et~al.}{2001}]{ADM01}
{Afanasiev} V.~L.,  {Dodonov} S.~N.,    {Moiseev} A.~V.,  2001, in {Ossipkov}
  L.~P.,  {Nikiforov} I.~I.,  eds, Stellar Dynamics: from Classic to Modern, 103

\bibitem[\protect\citeauthoryear{{Barazza}, {Binggeli} \& {Jerjen}}{{Barazza}
  et~al.}{2002}]{BBJ02}
{Barazza} F.~D.,  {Binggeli} B.,    {Jerjen} H.,  2002, \aap, 391, 823

\bibitem[\protect\citeauthoryear{{Bender} \& {Nieto}}{{Bender} \&
  {Nieto}}{1990}]{BN90}
{Bender} R.,  {Nieto} J.-L.,  1990, \aap, 239, 97

\bibitem[\protect\citeauthoryear{{Bertola} \& {Capaccioli}}{{Bertola} \&
  {Capaccioli}}{1975}]{BC75}
{Bertola} F.,  {Capaccioli} M.,  1975, \apj, 200, 439

\bibitem[\protect\citeauthoryear{{Binggeli}, {Sandage} \& {Tammann}}{{Binggeli}
  et~al.}{1985}]{BST85}
{Binggeli} B.,  {Sandage} A.,    {Tammann} G.~A.,  1985, \aj, 90, 1681

\bibitem[\protect\citeauthoryear{{Boselli}, {Boissier}, {Cortese} \&
  {Gavazzi}}{{Boselli} et~al.}{2008a}]{BBCG08_1}
{Boselli} A.,  {Boissier} S.,  {Cortese} L.,    {Gavazzi} G.,  2008, \apj, 674, 742

\bibitem[\protect\citeauthoryear{{Boselli}, {Boissier}, {Cortese} \&
  {Gavazzi}}{{Boselli} et~al.}{2008b}]{BBCG08}
{Boselli} A.,  {Boissier} S.,  {Cortese} L.,    {Gavazzi} G.,  2008, \aap, 489, 1015

\bibitem[\protect\citeauthoryear{{Cappellari} \& {Copin}}{{Cappellari} \&
  {Copin}}{2003}]{CC03}
{Cappellari} M.,  {Copin} Y.,  2003, \mnras, 342, 345

\bibitem[\protect\citeauthoryear{{Cappellari} \& {Emsellem}}{{Cappellari} \&
  {Emsellem}}{2004}]{CE04}
{Cappellari} M.,  {Emsellem} E.,  2004, \pasp, 116, 138

\bibitem[\protect\citeauthoryear{{Chilingarian}}{{Chilingarian}}{2006}]{Chilingarian06}
{Chilingarian} I.,  2006, PhD thesis, Moscow State University, Moscow \& Universit\'e
  Claude Bernard, Lyon, astro-ph/0611893

\bibitem[\protect\citeauthoryear{{Chilingarian}, {Cayatte} \&
  {Bergond}}{{Chilingarian} et~al.}{2008a}]{CCB08}
{Chilingarian} I.,  {Cayatte} V.,    {Bergond} G.,  2008a, MNRAS, 390, 906

\bibitem[\protect\citeauthoryear{{Chilingarian}, {Afanasiev}, {Bonnarel},
  {Dodonov}, {Louys} \& {Zolotukhin}}{{Chilingarian} et~al.}{2007a}]{ASPIDSR}
{Chilingarian} I., {Afanasiev} V., {Bonnarel} F., {Dodonov} S., {Louys} M., {Zolotukhin} I.
2007a, in {Shaw} R.~A., {Hill} F., {Bell} D.~J., eds, Astronomical Data Analysis Software and Systems XVI,
Vol.~376 of Astronomical Society of the Pacific Conference Series, 217

\bibitem[\protect\citeauthoryear{{Chilingarian}, {Prugniel}, {Sil'chenko} \&
  {Koleva}}{{Chilingarian} et~al.}{2007b}]{CPSK07}
{Chilingarian} I.,  {Prugniel} P.,  {Sil'chenko} O.,    {Koleva} M.,  2007b, in
  {Vazdekis} A.,  R.~{Peletier} R.,  eds, Stellar Populations as Building
  Blocks of Galaxies Vol.~241 of IAU Symposium,
Cambridge University Press, Cambridge, UK, 175

\bibitem[\protect\citeauthoryear{{Chilingarian}, {Sil'Chenko}, {Afanasiev} \&
  {Prugniel}}{{Chilingarian} et~al.}{2008b}]{CSAP08}
{Chilingarian} I.,  {Sil'Chenko} O.,  {Afanasiev} V.,    {Prugniel} P.,  2008b,
  in {Knapen} J.~H.,  {Mahoney} T.~J.,   {Vazdekis} A.,  eds, Pathways Through
  an Eclectic Universe Vol.~390 of Astronomical Society of the Pacific
  Conference Series, p.296

\bibitem[\protect\citeauthoryear{{Chilingarian}, {Cayatte}, {Durret}, {Adami},
  {Balkowski}, {Chemin}, {Lagan{\'a}} \& {Prugniel}}{{Chilingarian}
  et~al.}{2008c}]{Chilingarian+08}
{Chilingarian} I.~V.,  {Cayatte} V.,  {Durret} F.,  {Adami} C.,  {Balkowski}
  C.,  {Chemin} L.,  {Lagan{\'a}} T.~F.,    {Prugniel} P.,  2008c, \aap, 486, 85

\bibitem[\protect\citeauthoryear{{Chilingarian}, {Prugniel}, {Sil'chenko} \&
  {Afanasiev}}{{Chilingarian} et~al.}{2007c}]{CPSA07}
{Chilingarian} I.~V.,  {Prugniel} P.,  {Sil'chenko} O.~K.,    {Afanasiev}
  V.~L.,  2007c, \mnras, 376, 1033

\bibitem[\protect\citeauthoryear{{Chilingarian}, {Sil'chenko}, {Afanasiev} \&
  {Prugniel}}{{Chilingarian} et~al.}{2007d}]{CSAP07}
{Chilingarian} I.~V.,  {Sil'chenko} O.~K.,  {Afanasiev} V.~L.,    {Prugniel}
  P.,  2007d, Astronomy Letters, 33, 292

\bibitem[\protect\citeauthoryear{{Cid Fernandes}, {Mateus}, {Sodr{\'e}},
  {Stasi{\'n}ska} \& {Gomes}}{{Cid Fernandes} et~al.}{2005}]{CidFernandes+05}
{Cid Fernandes} R.,  {Mateus} A.,  {Sodr{\'e}} L.,  {Stasi{\'n}ska} G.,
  {Gomes} J.~M.,  2005, \mnras, 358, 363

\bibitem[\protect\citeauthoryear{{De Rijcke}, {Dejonghe}, {Zeilinger} \&
  {Hau}}{{De Rijcke} et~al.}{2001}]{DRDZH01}
{De Rijcke} S.,  {Dejonghe} H.,  {Zeilinger} W.~W.,    {Hau} G.~K.~T.,  2001,
  \apjl, 559, L21

\bibitem[\protect\citeauthoryear{{De Rijcke}, {Dejonghe}, {Zeilinger} \&
  {Hau}}{{De Rijcke} et~al.}{2003}]{DRDZH03}
{De Rijcke} S.,  {Dejonghe} H.,  {Zeilinger} W.~W.,    {Hau} G.~K.~T.,  2003,
  \aap, 400, 119

\bibitem[\protect\citeauthoryear{{De Rijcke}, {Dejonghe}, {Zeilinger} \&
  {Hau}}{{De Rijcke} et~al.}{2004}]{DRDZH04}
{De Rijcke} S.,  {Dejonghe} H.,  {Zeilinger} W.~W.,    {Hau} G.~K.~T.,  2004,
  \aap, 426, 53

\bibitem[\protect\citeauthoryear{{De Rijcke}, {Michielsen}, {Dejonghe},
  {Zeilinger} \& {Hau}}{{De Rijcke} et~al.}{2005}]{deRijcke+05}
{De Rijcke} S.,  {Michielsen} D.,  {Dejonghe} H.,  {Zeilinger} W.~W.,    {Hau}
  G.~K.~T.,  2005, \aap, 438, 491

\bibitem[\protect\citeauthoryear{{Ferguson} \& {Binggeli}}{{Ferguson} \&
  {Binggeli}}{1994}]{FB94}
{Ferguson} H.~C.,  {Binggeli} B.,  1994, \aapr, 6, 67

\bibitem[\protect\citeauthoryear{{Ferrarese}, {C{\^o}t{\'e}}, {Jord{\'a}n},
  {Peng}, {Blakeslee}, {Piatek}, {Mei}, {Merritt}, {Milosavljevi{\'c}}, {Tonry}
  \& {West}}{{Ferrarese} et~al.}{2006}]{Ferrarese+06}
{Ferrarese} L.,  {C{\^o}t{\'e}} P.,  {Jord{\'a}n} A.,  {Peng} E.~W.,
  {Blakeslee} J.~P.,  {Piatek} S.,  {Mei} S.,  {Merritt} D.,
  {Milosavljevi{\'c}} M.,  {Tonry} J.~L.,    {West} M.~J.,  2006, \apjs, 164,
  334

\bibitem[\protect\citeauthoryear{{Fioc} \& {Rocca-Volmerange}}{{Fioc} \&
  {Rocca-Volmerange}}{1997}]{FR97}
{Fioc} M.,  {Rocca-Volmerange} B.,  1997, \aap, 326, 950

\bibitem[\protect\citeauthoryear{{Fukugita}, {Shimasaku} \&
  {Ichikawa}}{{Fukugita} et~al.}{1995}]{FSI95}
{Fukugita} M.,  {Shimasaku} K.,    {Ichikawa} T.,  1995, \pasp, 107, 945

\bibitem[\protect\citeauthoryear{{Geha}, {Guhathakurta} \& {van der
  Marel}}{{Geha} et~al.}{2002}]{GGvdM02}
{Geha} M.,  {Guhathakurta} P.,    {van der Marel} R.~P.,  2002, \aj, 124, 3073

\bibitem[\protect\citeauthoryear{{Geha}, {Guhathakurta} \& {van der
  Marel}}{{Geha} et~al.}{2003}]{GGvdM03}
{Geha} M.,  {Guhathakurta} P.,    {van der Marel} R.~P.,  2003, \aj, 126, 1794

\bibitem[\protect\citeauthoryear{{Geha}, {Guhathakurta} \& {van der
  Marel}}{{Geha} et~al.}{2005}]{GGvdM05}
{Geha} M.,  {Guhathakurta} P.,    {van der Marel} R.~P.,  2005, \aj, 129, 2617

\bibitem[\protect\citeauthoryear{{Graham} \& {Guzm{\'a}n}}{{Graham} \&
  {Guzm{\'a}n}}{2003}]{GG03}
{Graham} A.~W.,  {Guzm{\'a}n} R.,  2003, \aj, 125, 2936

\bibitem[\protect\citeauthoryear{{Gunn} et~al.}{{Gunn} et~al.}{2006}]{Gunn+06}
{Gunn} J.~E. et~al., 2006, \aj, 131, 2332

\bibitem[\protect\citeauthoryear{{Jerjen}, {Kalnajs} \& {Binggeli}}{{Jerjen}
  et~al.}{2000}]{JKB00}
{Jerjen} H.,  {Kalnajs} A.,    {Binggeli} B.,  2000, \aap, 358, 845

\bibitem[\protect\citeauthoryear{{Janz} \& {Lisker}}{{Janz} \&
  {Lisker}}{2008}]{JL08}
{Janz} J.,  {Lisker} T.,  2008, \apjl, 689, L25

\bibitem[\protect\citeauthoryear{{Koleva}, {Bavouzet}, {Chilingarian} \&
  {Prugniel}}{{Koleva} et~al.}{2007}]{KBCP07}
{Koleva} M.,  {Bavouzet} N.,  {Chilingarian} I.,    {Prugniel} P.,  2007, in
  {Kissler-Patig} M.,  {Walsh} J.~R.,   {Roth} M.~M.,  eds, Science
  Perspectives for 3D Spectroscopy, 153

\bibitem[\protect\citeauthoryear{{Koleva}, {Prugniel}, {Ocvirk}, {Le Borgne} \&
  {Soubiran}}{{Koleva} et~al.}{2008}]{Koleva+08}
{Koleva} M.,  {Prugniel} P.,  {Ocvirk} P.,  {Le Borgne} D.,    {Soubiran} C.,
  2008, \mnras, 385, 1998

\bibitem[\protect\citeauthoryear{{Kormendy}, {Fisher}, {Cornell} \&
  {Bender}}{{Kormendy} et~al.}{2008}]{KFCB08}
{Kormendy} J.,  {Fisher} D.~B.,  {Cornell} M.~E.,    {Bender} R.,  2008,
  Accepted to ApJS, ArXiv:0810.1681

\bibitem[\protect\citeauthoryear{{Kuntschner}}{{Kuntschner}}{2004}]{Kuntschner%
04}
{Kuntschner} H.,  2004, \aap, 426, 737

\bibitem[\protect\citeauthoryear{{Kuntschner}, {Emsellem}, {Bacon}, {Bureau},
  {Cappellari}, {Davies}, {de Zeeuw}, {Falc{\'o}n-Barroso}, {Krajnovi{\'c}},
  {McDermid}, {Peletier} \& {Sarzi}}{{Kuntschner} et~al.}{2006}]{Kuntschner+06}
{Kuntschner} H.,  {Emsellem} E.,  {Bacon} R.,  {Bureau} M.,  {Cappellari} M.,
  {Davies} R.~L.,  {de Zeeuw} P.~T.,  {Falc{\'o}n-Barroso} J.,  {Krajnovi{\'c}}
  D.,  {McDermid} R.~M.,  {Peletier} R.~F.,    {Sarzi} M.,  2006, \mnras, 369,
  497

\bibitem[\protect\citeauthoryear{{Le Borgne}, {Rocca-Volmerange}, {Prugniel},
  {Lan{\c c}on}, {Fioc} \& {Soubiran}}{{Le Borgne} et~al.}{2004}]{LeBorgne+04}
{Le Borgne} D.,  {Rocca-Volmerange} B.,  {Prugniel} P.,  {Lan{\c c}on} A.,
  {Fioc} M.,    {Soubiran} C.,  2004, \aap, 425, 881

\bibitem[\protect\citeauthoryear{{Lisker}, {Grebel} \& {Binggeli}}{{Lisker}
  et~al.}{2006}]{LGB06}
{Lisker} T.,  {Grebel} E.~K.,    {Binggeli} B.,  2006, \aj, 132, 497

\bibitem[\protect\citeauthoryear{{Lisker}, {Grebel} \& {Binggeli}}{{Lisker}
  et~al.}{2008}]{LGB08}
{Lisker} T.,  {Grebel} E.~K.,    {Binggeli} B.,  2008, \aj, 135, 380

\bibitem[\protect\citeauthoryear{{Michielsen}, {Koleva}, {Prugniel},
  {Zeilinger}, {De Rijcke}, {Dejonghe}, {Pasquali}, {Ferreras} \&
  {Debattista}}{{Michielsen} et~al.}{2007}]{Michielsen+07}
{Michielsen} D.,  {Koleva} M.,  {Prugniel} P.,  {Zeilinger} W.~W.,  {De Rijcke}
  S.,  {Dejonghe} H.,  {Pasquali} A.,  {Ferreras} I.,    {Debattista} V.~P.,
  2007, \apjl, 670, L101

\bibitem[\protect\citeauthoryear{{Michielsen}, {Boselli}, {Conselice}, 
{Toloba}, {Whiley}, {Arag{\'o}n-Salamanca}, {Balcells},{Cardiel}, 
{Cenarro}, {Gorgas}, {Peletier} \& {Vazdekis}}{{Michielsen} 
et~al.}{2008}]{Michielsen+08}
{Michielsen} D., {Boselli} A., {Conselice} C.~J.,
{Toloba} E., {Whiley} I.~M., {Arag{\'o}n-Salamanca} A.,
{Balcells} M., {Cardiel} N., {Cenarro} A.~J., {Gorgas} J.
{Peletier} R.~F., {Vazdekis} A., 2008, \mnras, 385, 1374

\bibitem[\protect\citeauthoryear{{Moore}, {Katz}, {Lake}, {Dressler} \&
  {Oemler}}{{Moore} et~al.}{1996}]{Moore+96}
{Moore} B.,  {Katz} N.,  {Lake} G.,  {Dressler} A.,    {Oemler} A.,  1996,
  \nat, 379, 613

\bibitem[\protect\citeauthoryear{{Ocvirk}, {Pichon}, {Lan{\c c}on} \&
  {Thi{\'e}baut}}{{Ocvirk} et~al.}{2006}]{OPLT06}
{Ocvirk} P.,  {Pichon} C.,  {Lan{\c c}on} A.,    {Thi{\'e}baut} E.,  2006,
  \mnras, 365, 74

\bibitem[\protect\citeauthoryear{{Paturel}, {Petit}, {Prugniel}, {Theureau},
  {Rousseau}, {Brouty}, {Dubois} \& {Cambr{\'e}sy}}{{Paturel}
  et~al.}{2003}]{Paturel+03}
{Paturel} G.,  {Petit} C.,  {Prugniel} P.,  {Theureau} G.,  {Rousseau} J.,
  {Brouty} M.,  {Dubois} P.,    {Cambr{\'e}sy} L.,  2003, \aap, 412, 45

\bibitem[\protect\citeauthoryear{{Pedraz}, {Gorgas}, {Cardiel},
  {S{\'a}nchez-Bl{\'a}zquez} \& {Guzm{\'a}n}}{{Pedraz}
  et~al.}{2002}]{Pedraz+02}
{Pedraz} S.,  {Gorgas} J.,  {Cardiel} N.,  {S{\'a}nchez-Bl{\'a}zquez} P.,
  {Guzm{\'a}n} R.,  2002, \mnras, 332, L59

\bibitem[\protect\citeauthoryear{{Prugniel}, {Chilingarian}, {Sil'Chenko} \&
  {Afanasiev}}{{Prugniel} et~al.}{2005}]{PCSA05}
{Prugniel} P.,  {Chilingarian} I.,  {Sil'Chenko} O.,    {Afanasiev} V.,  2005,
  in {Jerjen} H.,  {Binggeli} B.,  eds, IAU Colloq. 198: Near-fields cosmology
  with dwarf elliptical galaxies, 73

\bibitem[\protect\citeauthoryear{{Prugniel}, {Soubiran}, {Koleva} \& {Le
  Borgne}}{{Prugniel} et~al.}{2007}]{PSKLB07}
{Prugniel} P.,  {Soubiran} C.,  {Koleva} M.,    {Le Borgne} D.,  2007, astro-ph/0703658

\bibitem[\protect\citeauthoryear{{Rakos}, {Schombert}, {Maitzen}, {Prugovecki}
  \& {Odell}}{{Rakos} et~al.}{2001}]{Rakos+01}
{Rakos} K.,  {Schombert} J.,  {Maitzen} H.~M.,  {Prugovecki} S.,    {Odell} A.,
   2001, \aj, 121, 1974

\bibitem[\protect\citeauthoryear{{Salpeter}}{{Salpeter}}{1955}]{Salpeter55}
{Salpeter} E.~E.,  1955, \apj, 121, 161

\bibitem[\protect\citeauthoryear{{Sil'chenko}}{{Sil'chenko}}{2006}]{Sil06}
{Sil'chenko} O.,  2006, \apj, 641, 229

\bibitem[\protect\citeauthoryear{{Simien} \& {Prugniel}}{{Simien} \&
  {Prugniel}}{2002}]{SP02}
{Simien} F.,  {Prugniel} P.,  2002, \aap, 384, 371

\bibitem[\protect\citeauthoryear{{Stiavelli}, {Miller}, {Ferguson}, {Mack},
  {Whitmore} \& {Lotz}}{{Stiavelli} et~al.}{2001}]{Stiavelli+01}
{Stiavelli} M.,  {Miller} B.~W.,  {Ferguson} H.~C.,  {Mack} J.,  {Whitmore}
  B.~C.,    {Lotz} J.~M.,  2001, \aj, 121, 1385

\bibitem[\protect\citeauthoryear{{Thomas}, {Brimioulle}, {Bender}, {Hopp},
  {Greggio}, {Maraston} \& {Saglia}}{{Thomas} et~al.}{2006}]{Thomas+06}
{Thomas} D.,  {Brimioulle} F.,  {Bender} R.,  {Hopp} U.,  {Greggio} L.,
  {Maraston} C.,    {Saglia} R.~P.,  2006, \aap, 445, L19

\bibitem[\protect\citeauthoryear{{Thomas}, {Maraston} \& {Bender}}{{Thomas}
  et~al.}{2003}]{TMB03}
{Thomas} D.,  {Maraston} C.,    {Bender} R.,  2003, \mnras, 339, 897

\bibitem[\protect\citeauthoryear{{van der Marel} \& {Franx}}{{van der Marel} \&
  {Franx}}{1993}]{vdMF93}
{van der Marel} R.~P.,  {Franx} M.,  1993, \apj, 407, 525

\bibitem[\protect\citeauthoryear{{van Dokkum}}{{van
  Dokkum}}{2001}]{vanDokkum01}
{van Dokkum} P.~G.,  2001, \pasp, 113, 1420

\bibitem[\protect\citeauthoryear{{van Zee}, {Barton} \& {Skillman}}{{van Zee}
  et~al.}{2004a}]{vZBS04}
{van Zee} L.,  {Barton} E.~J.,    {Skillman} E.~D.,  2004a, \aj, 128, 2797

\bibitem[\protect\citeauthoryear{{van Zee}, {Skillman} \& {Haynes}}{{van Zee}
  et~al.}{2004b}]{vZSH04}
{van Zee} L.,  {Skillman} E.~D.,    {Haynes} M.~P.,  2004b, \aj, 128, 121

\bibitem[\protect\citeauthoryear{{White} \& {Frenk}}{{White} \&
  {Frenk}}{1991}]{WF91}
{White} S.~D.~M.,  {Frenk} C.~S.,  1991, \apj, 379, 52

\bibitem[\protect\citeauthoryear{{Worthey}}{{Worthey}}{1994}]{Worthey94}
{Worthey} G.,  1994, \apjs, 95, 107

\bibitem[\protect\citeauthoryear{{Worthey}, {Faber}, {Gonzalez} \&
  {Burstein}}{{Worthey} et~al.}{1994}]{WFGB94}
{Worthey} G.,  {Faber} S.~M.,  {Gonzalez} J.~J.,    {Burstein} D.,  1994,
  \apjs, 94, 687

\end{thebibliography}

\label{lastpage}

\end{document}